\documentclass[12pt]{article}
\usepackage{amsfonts}
\usepackage{eurosym}
\usepackage{graphicx}
\usepackage{booktabs}
\usepackage{float}
\usepackage{subfig}
\usepackage{amssymb}
\usepackage{comment}
\usepackage{amsmath}
\usepackage{booktabs}
\usepackage{multirow}
\usepackage{enumerate}
\usepackage{color}
\usepackage{xcolor}
\usepackage[authoryear]{natbib}
\usepackage{hyperref}

\newtheorem{lemma}{{\bf \sc Lemma}}

\newtheorem{proposition}{{\bf \sc Proposition}}

\usepackage{hyperref}
\usepackage{cleveref}
\hypersetup{colorlinks=false,linkcolor={blue},citecolor={blue},urlcolor={red}}
\usepackage[title]{appendix}
\usepackage{setspace}
\usepackage{layouts}

\newcommand\epigraph[2]{
\hfill{}\begin{minipage}{5.8in}{\begin{spacing}{0.9}
\noindent\textit{#1}\end{spacing}
\hfill{}{#2}}\vspace{1em}
\end{minipage}}

\def\eproof{\hbox{\hskip3pt\vrule width4pt height8pt depth1.5pt}}

\allowdisplaybreaks

\begin{document}

\title{ Learning through the Grapevine: The Impact of Noise and the Breadth and Depth of Social Networks}
\author{Matthew O. Jackson, Suraj Malladi, and David McAdams \thanks{%
		Jackson is from the Department of Economics, Stanford University, Stanford, California 94305-6072 USA,
		and is also an external faculty member at the Santa Fe Institute.
		Malladi is from the Graduate School of Business, Stanford University.
		McAdams is at the Fuqua School of Business and Economics Department, Duke University.
		Emails: jacksonm@stanford.edu, surajm@stanford.edu, david.mcadams@duke.edu.
		We gratefully acknowledge financial support under NSF grant SES-1629446 and from Microsoft Research New England.
		We thank Arun Chandrasekhar, Ben Golub, David Hirshleifer, Mallesh Pai, Evan Sadler, and Omer Tamuz for helpful conversations and suggestions. }}
\date{Draft: June 2020}
\maketitle

\begin{abstract}
We examine how well people learn when information is noisily relayed from person to person;
and we study how communication platforms can improve learning without
censoring or fact-checking messages.
We analyze learning as a function of social network depth
(how many times information is relayed) and breadth (the number of relay chains accessed).
Noise builds up as depth increases, so learning requires greater breadth.
In the presence of mutations (deliberate or random) and transmission failures of messages,
we characterize sharp thresholds for breadths above which receivers learn fully and below which they
learn nothing. When there is uncertainty about mutation rates,
optimizing learning requires either capping depth, or
if that is not possible, limiting breadth by capping the number of people
to whom someone can forward a message.
Limiting breadth cuts the number of messages received but also decreases the fraction originating
further from the receiver, and so can increase the signal to noise ratio.
Finally, we extend our model to study learning from message survival: e.g., people are more likely to
pass messages with one conclusion than another.
We find that as depth grows, all learning comes from either the total number of messages
received or from the content of received messages, but the learner does not need to pay attention to both.


\end{abstract}

\thispagestyle{empty}

\setcounter{page}{0} \newpage

\epigraph{The safety of our democracy is more important than shareholder dividends and CEO salaries, and we need tech companies to behave accordingly. That’s why I'm calling on them to take real steps right now to fight disinformation.}{Elizabeth Warren}   %https://elizabethwarren.com/plans/fighting-digital-disinformation

\epigraph{I don't think that Facebook or internet platforms in general should be arbiters of truth.}{Mark Zuckerberg}   %https://www.cnbc.com/2020/05/28/zuckerberg-facebook-twitter-should-not-fact-check-political-speech.html

\section{Introduction}

People rely on word-of-mouth learning when deciding
whether to vaccinate their children, adopt a new diet, participate in an aid program, adopt a new technology, or support
some government policy.
In some cases, such learning fails to deliver a consensus on basic facts that are a vital basis for such
decisions, even when there is consensus among primary information sources.
For example, \cite{largent2012vaccine} estimates that 40 percent
of American parents have delayed or denied recommended vaccinations
for their children.  Also, only around 73 percent of Americans
believe that climate change is happening, and even fewer believe
that humans have had an impact on the climate.\footnote{This figure
is from the December 2018 ``Climate Change in the American Mind'' poll
conducted jointly by the Yale Program on Climate Change Communication
and the George Mason University Center for Climate Change Communication.}
These disparities reflect, in part, the fact that communication is subject to substantial bias and noise.
Word-of-mouth information can be relayed through long chains -- especially on modern platforms.  People hear
many things, but what they hear may no longer reflect the original scientific evidence, either because it is improperly relayed, deliberately distorted, or selectively
dropped along the way.
When does noise render learning impossible? Can platforms improve learning without verifying and
policing content relayed by users?%, but instead simply by controlling the number of times a message can be relayed?

We investigate how people learn as a function of their networks when information is subject to mutation, deliberate manipulation, and content-based dropping.
In particular, we characterize how learning depends on the depth and breadth of a person's network.
Since information that travels a longer path is less likely to survive and be accurate if it does survive, increasing network
size does not necessarily improve learning.
Both depth and breadth increase the size of a network, but they also increase the relative amount of
noise by increasing the relative number of sources at greater distances compared to those closer by.
We examine how depth and breadth impact learning and how their effects relate to each other.
Our analysis shows how limiting the network can improve the accuracy of overall content, without the need for private or public monitoring of messages or censorship.
This is important, given the many problems associated with censorship, even with benign intentions. Our
results show how one can satisfy Warren's objectives outlined in
the quote above, while respecting
Zuckerberg's reticence for involving private enterprise in censorship.

In our model, information is
relayed from original sources via sequences of individuals to an
eventual receiver, who wishes to learn the state of the world. We code the
state of the world and messages as either being in favor of an action (``1'', e.g., it is best to vaccinate a child) or against it (``0'', it is best not to vaccinate). With noiseless word-of-mouth communication and sufficiently many starting sources of conditionally independent information, the learner learns the true state. However, along each chain, the message may mutate or be dropped -- reducing the information content of the signals that reach the learner.

First, in a world in which the learner knows mutation and dropping rates,  we show that a person learns the state (with a probability approaching 1 as depth increases) if and only if the number of chains that they have
access to (the breadth of their network) exceeds a threshold that grows exponentially in the depth of the network, as well as in
the mutation and dropping rates.
Basically, the increased noise from greater depth has to be countered by more total sources.
This provides a precise relationship between breadth and depth: breadth has to exceed a threshold that
increases with depth (or equivalently, depth has to be lower than a threshold that increases with breadth) in order to make learning possible.   The threshold is sharp in that full learning happens asymptotically above the threshold and no learning below the threshold.

Next, we show that small amounts of uncertainty about mutation rates precludes learning from long chains.
The intuition here is that over long chains most messages that survive have mutated, and in the limit,
information content disappears.
The ratio of 1's to 0's is slightly different depending on the starting state,
but that difference vanishes as chains get longer.
If there is any uncertainty about the relative likelihood of mutating from 1 to 0 or vice versa,
then that uncertainty swamps the tiny differences
that emerge from the starting state.  Learning is completely precluded.

The key to overcoming this is to limit the depth of messages, or at least the relative ratio of messages that are coming from far away.
This is the subject of some of our key results.
By limiting the total distance that messages can travel, one
limits the chances that messages are distorted. This allows for partial learning from nearby
messages.   People see fewer messages, but
ones that are more likely to be informative - and this increases the overall
signal to noise ratio.
Thus, capping depth can be helpful, and we characterize an optimal cap.
We also explore what happens when depth is not easily capped.  For instance, some social media platforms do no track whether a message is new or somehow forwarded.
However, it is easy for them to track how many people someone broadcasts a message to.\footnote{Users
might individually re-target messages, but this becomes much more burdensome.}
By capping the number of people a person can send a message to
(or at least making it more difficult), one
can control the breadth of the network.\footnote{What matters is average in-degree,
but what can be most easily capped is out-degree.  However, note that average in-degree must equal average
out-degree, and so capping one caps the other.}
This can also help, since decreasing breadth increases the {\sl relative} number of nodes in a
network that are close compared to farther away for any given depth.
Thus, without being able to control depth, limiting breadth can also improve learning.
Again, we provide bounds and explore optimal policies.

Interestingly, breadth-limits have been adopted by online messaging platforms.
For instance, {\sl WhatsApp} has capped the number of people that someone can message for the express purpose of curbing the spread of false
information.

Finally, we extend the model to study how well a receiver learns when dropping rates also depend on the message being relayed.
For instance, a person may be more likely to pass along information that they find surprising,
or that is in line with their prior beliefs.\footnote{In \cite{banerjee1993}, a rumor's survival rate increases with the number
	of people who have made an investment. Hearing a rumor therefore conveys information about how
	many people invested.  The survival bias here comes from a different source, but there is still
	an inference to be made purely from information survival regardless of content.}
Even in the publication of scientific articles,
reviewers may be more likely to agree to publish (pass along) statistically significant
or surprising results than insignificant or expected ones.
In this case, a receiver can learn from how many messages she receives.
Hearing about very few studies, even though many were conducted, is informative:
the studies most likely did not find `exciting' results
that prompted them to be discussed and forwarded.
For example, if 0's are more likely to be dropped than 1's, then significantly more messages make
it to the learner when the true state is 1.  Thus,
even without looking at the content, a learner can infer something about the state simply
by tracking how many messages reach her.

We bound how much more likely a fully Bayesian agent -- who
updates based on both message survival and content -- is to guess the state compared to someone who
looks only at message content or only at message survival. We show that the Bayesian's advantage
vanishes as distance to primary sources increases: for any parameters of the model, all the information
is contained in either message frequency alone or in message content alone.
This implies that full Bayesian learning is fully approximated in the limit by simple rules of
thumb conditioned on just one dimension of the available information.  Thus, learning need not involve
sophisticated calculations but simple thresholds:
believe 1 if and only if more than a certain number of messages are received, or if and only if the ratio of messages containing 1's is above some threshold.

\paragraph{Some Background Observations}

To motivate our modeling of long chains of messages that involve mutations and droppings,
we first discuss some accounts of noisy communication and the challenges of aggregating decentralized
information.

There is reason to believe that humans can learn well when they have access to each other's knowledge.
For instance, Francis Galton's famous article ``Vox Populi'' \citeyearpar{galton1907} showed that  the information possessed by a group of people, when centrally aggregated, can be remarkably accurate.
Galton examined 787 entries in a contest at the ``West of England Fat Stock and Poultry Fair,''
in which people guessed the dressed weight of an ox. Even though
more than half of the guesses were off by more than 3 percent,
the mean and median guesses were each within 0.1 percent of the ox's
true weight.
This dispersed information can hypothetically be correctly aggregated in a
decentralized manner through repeated communication in a social network.\footnote%
{\cite{golubj2010} show that individuals can converge to accurate beliefs by repeatedly (weightedly) averaging their beliefs with those of their neighbors, as long as the social network is balanced in a sense that no individual is unduly influential. \cite*{mueller2014} shows that, if some individuals are more sophisticated, then learning can occur in a broader set of networks.
For a review of the large literature on social learning that we do not survey here; see e.g., \cite{golubs2016}.}

Although such aggregation is feasible via relayed messages in a network,
there remain many important situations in which people fail to reach a consensus, and thus at least some
of them do not learn the truth, as evidenced by the disparity in beliefs regarding vaccination and climate change, cited above.
Relayed information is subject to substantial noise, which can be a barrier.
Consider the children's game of `Telephone', in which a starting message is whispered
from one player to the next. The final message typically bears little resemblance to the original message because of ``mutations'' that occur along the transmission chain.
Such mutations can and do happen in many environments.  \cite*{simmonsetal2011} discuss a revealing example of message mutation.  An initial tweet,
``Street style shooting in Oxford Circus for ASOS and Diet Coke. Let me know if you're around!'', was an invitation for people to join the crowd for a commercial being filmed in London. This was misunderstood and within minutes had mutated to ``Shooting in progress in Oxford Circus? What?'' and then retweeted as ``Shooting in progress in Oxford Circus, stay safe people.'' The informational content of the message completely changed.

Such mutations have been found to occur frequently. In \cite*{adamicetal2016}'s study of online viral memes,
one meme was reposted more than 470,000 times, with a mutation rate of around 11 percent and
more than 100,000 variants. This was not an outlier in their analysis: 121 of the 123 most viral
memes each had more than 100,000 variants.\footnote{Other examples of mutating messages include mythology and the morphing of religious texts.
\cite*{gurry2016number} estimates that there are around 500,000 textual variants of the Greek New Testament, not including spelling errors.}

Message mutations need not be symmetric communication errors. For example, messages may often drift to become more exaggerated as they are passed along, while exaggerated messages are less likely to be conveyed in a more muted form.
Message mutations need not be unintentional either.
Another source of asymmetries in mutations can be the presence of ideologues who
relay messages that they prefer telling rather than what
they heard (e.g., ``fake news''). There may be more ideologues pushing for one version of the facts
than there are for another. Messages may accordingly drift more frequently into one telling
than out of
it.\footnote{For other perspectives on the role of biased agents in the spread of
false information, see \cite*{acemoglu2010spread} and
\cite*{bloch2018rumors}. }

Indeed, ideologically-charged agents who manipulate information, or who are suspected
to manipulate information, may seriously hinder learning.
To give an example, \cite{largent2012vaccine} finds
that parents who are skeptical of vaccinations have ``tremendously high
trust in medical communities.'' But he writes ``[who] don't they trust?
The feds, and pharma.''\footnote{This was quoted from Mark Largent's book,
\textit{Vaccine: The Debate in Modern America}, in \textit{The Atlantic}
article ``Anti-Vaxers Aren't Stupid'' by Emma Green.} These are agents
who participate in the process of diffusing scientific research and whom
some parents worry may be ideologically motivated to misreport scientific
research.

All these types of distortions -- random mutations, exaggeration or bias -- build up as relaying chains grow.
And with online channels in particular, the paths that word-of-mouth communication follow can be quite
long. \cite{liben2008tracing} found instances of Internet
chain letters that traveled median
distances of over one hundred links.\footnote{ \cite{golubj2010b} explain why the resulting trees can be
much longer than they are wide.}  \cite{adamicetal2016} examined hundreds of millions of instances of thousands of memes and found chains with lengths in the hundreds and typical distances well into the dozens.

This potential for facilitating noise has certainly been realized on large messaging platforms and
online social networks like Twitter, Facebook, and WhatsApp.  Several governments have increased regulations
and fines for disinformation (\cite{posner_2019}).% (``dis''-information is deliberate).
The responsibility to curb misinformation and disinformation puts platforms
that are hesitant to be ``arbiters of the truth'' in a difficult situation.
Our analysis highlights a policy that such platforms could follow, without needing to
monitor or police message content. Limiting message passing does not eliminate false information
but can improve the overall quality of what gets shared and allow people to learn more
from the content they encounter.

\section{The Base Model of Noisy Information Transmission}\label{model}

We begin by studying how noise builds up along a chain that can travel a path of length $T$ from an
original source to ``the learner.''

Information passes by ``word of mouth,''  but this can be oral, written, via social media, etc.

There are two possible states of the world, $\omega \in \{0, 1\}$,
and $\theta\in (0,1)$ is the prior probability that the state is 1.

A sequence of agents $\{1,2,\ldots, T\}$, referred to as a ``chain,'' successively relays a signal of
the state via word of mouth, terminating with the learner at $T\geq 1$.

We do not model what the learner does with this information, but one can think of
the learner preferring to match her action with the state.
For instance, the learner may hear from friends about whether there is a link between
vaccines and autism and then decide whether to vaccinate her child.

A first agent in a chain, interpreted as ``an original source,''
observes a noisy signal of the state, $s_1 \in \{0,1, \emptyset\}$.\footnote%
{We focus on a binary world to crystallize the main ideas.  Extensions to richer state spaces and signal structures are left for future research.}
That signal is transmitted with noise becoming $s_2\in \{0,1, \emptyset\}$, and so on
until signal $s_T$ reaches the learner.

The ``null signal,'' $s_{t} = \emptyset$, indicates that no signal was received,
in which case no signal is transmitted.  Another possibility is that something was received,
but that the information was sent along in some incoherent manner:  one person hears from another but cannot understand what was said and so has no information to pass along.
In particular, if agent $t\geq 1$ receives the null signal $s_{t} = \emptyset$,
then all subsequent agents (including the learner) also receive the null signal.

If agent $t\geq 1$ receives a signal $s_t \in \{0,1\}$,
then that agent passes a signal along ($s_{t+1}\neq \emptyset$)
with probability $p_1$ if $s_t = 1$, and with probability $p_0$ if $s_t = 0$.
Thus, for instance, if $p_1 > p_0$ then agents are more likely to transmit a signal if they heard a 1, and vice versa if $p_1 < p_0$.  With the remaining probabilities of $1-p_1$ and $1-p_0$, respectively, the signal is dropped and $s_{t+1}=\emptyset$.

Each time a non-null signal is transmitted, that signal mutates from 0 to 1 with probability
$\mu_{01}\in [0,1/2)$, or from 1 to 0 with probability $\mu_{10}\in [0,1/2)$; we let $M \equiv  1 - \mu_{01} - \mu_{10}$.
We presume that these mutation rates are less than 1/2 so that we do not end up with signals continuously reversing themselves.
Again, these mutations could be from a person deliberately changing a message to what their preferred signal would be, or could be due to some misunderstanding or other noise in the communication.

In summary, if $s_{t-1}=1$, the next agent (including $t=1$) hears $s_t=1$ with probability $p_1(1-\mu_{10})$, $s_t=0$ with probability $p_1\mu_{10}$, and $s_t=\emptyset$ with probability $1-p_1$.
Similarly, conditional on $s_{t-1}=0$, $s_t=1$ with
probability $p_0\mu_{01}$, $s_t=0$ with probability $p_0(1-\mu_{01})$,
and $s_t=\emptyset$ with probability $1-p_0$.
If $s_t=\emptyset$ for some $t$, then that is true for all subsequent signals.  This defines a $3\times 3$ Markov chain in which $\emptyset$ is an absorbing state.
\footnote{Note that the setting is stationary in that
the initial signal $s_1$ is derived from the original state in the same way as any other $s_t$ depends on $s_{t-1}$,
as if nature were ``agent 0'' in the chain with signal $s_0$ equal to the state.
Our analysis easily extends when
	first-signal accuracies and dropping rates differ from subsequent ones; but this assumption simplifies the expressions.}

\begin{figure}[h!]
	\centering
	\includegraphics[scale=0.5]{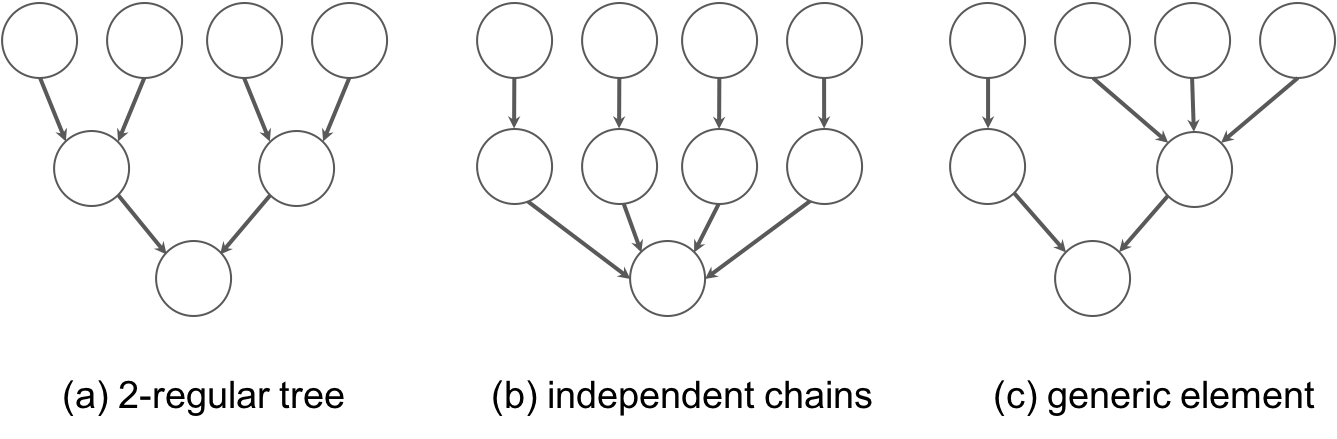}
	\caption{Three trees in $R(4, 2)$, the set of depth-2 directed trees with 4 leaves.}
	\label{fig:tree_examples}
\end{figure}

The first part of our analysis presumes that the learner has access to
some number $n \geq 1$ of length $t$ chains of messages, relayed through some social network.
This network is some depth $t$ directed tree of agents, with $n$ leaves representing the original sources, the root representing the learner, and edges representing the direction of relay. Each path from the leaves to the root is a chain along which messages are forwarded.
We let $R(n, t)$ be the set of such trees, with three examples pictured in Figure \ref{fig:tree_examples}.

\begin{figure}[h!]
	\centering
	\includegraphics[scale=0.4]{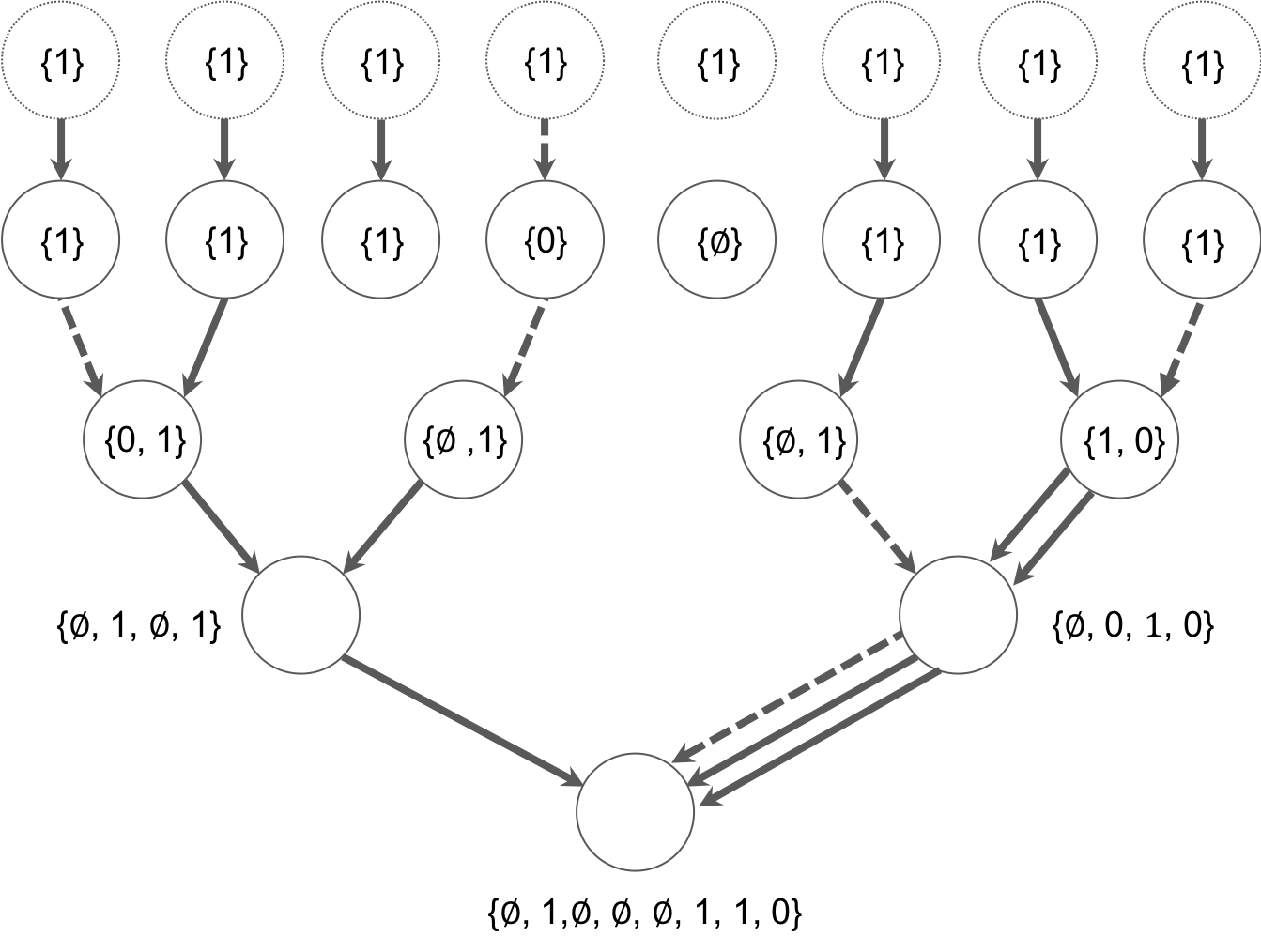}
	\caption{
The root node (``learner'') receives messages passed through eight paths, each starting from a different source. The absence of an arrow from one node to the one below it indicates that no message was sent, a dashed arrow indicates the message was delivered but mutated, and a solid arrow indicates that the message was delivered un-mutated. In this example, the true state is $1$ and paths 1-3 and 6-8 begin with a correct initial signal, while path 4 begins with an incorrect initial signal and path 5 begins with no signal received. Initial messages are delivered on paths 1,2,4, and 6-8, mutating from 1 to 0 on path 1 and from 0 to 1 on path 4, and undelivered on paths 3 and 5. Messages are then  relayed on path 2,4, and 6-8, mutating from 1 to 0 on path 6, but dropped on path 1. Finally, messages are re-relayed on paths 2 and 6-8, mutating from 0 to 1 on path 6, but dropped on path 4. Overall, the learner hears four messages, of which two never mutated, one mutated once, and one mutated twice.}
	\label{fig:comm_diagram}
\end{figure}

Conditionally independent signals of the state
are independently relayed along each of these chains of length $t$ via the same noisy process to the same learner.
An example of the communication process over a 2-regular, depth 4 tree is pictured in Figure \ref{fig:comm_diagram}.

\section{Learning from Chains of Noisy Transmission}\label{section_learning}

In this section we consider the case in which signals mutate, but their content does not affect the likelihood of dropping
($p_0 = p_1=p$).
This allows us to focus on how people update based on the content of the information that they receive, rather than whether they received information at all.
We return to discuss learning from content-dependent signal survival ($p_0 \neq p_1$) in Section \ref{subsection_survival}.

\subsection{Learning from Mutated Messages}\label{subsection_mutation}

We begin by establishing conditions for learning in any network in $R(n, T)$.

Suppose that the learner has access to $n(T)$ independent chains of length $T$.
We index $n$ by $T$ since we wish to characterize how many chains are needed as a function of their length.  Longer chains are more likely to be null or to have an incorrect signal and so more are needed to deliver an equivalent amount of information.
Let $I_{n(T)}$ be the vector of (potentially null) random signals that the learner receives from the chains, and let
the random variable $b(n(T),T)=\Pr_T (\omega=1 | I_{n(T)} )$ be the posterior probability that the state equals 1 conditional on the information from $n(T)$ originating signals that have each independently traveled $T$ steps.

We say that
$\tau(T)$  is a {\sl threshold for learning}  if (i) $Plim \ b(n(T),T) = 1$ or $0$ whenever $n(T)/\tau(T) \rightarrow \infty$ and (ii) $Plim \  b(n(T),T) = \theta$ whenever $n(T)/\tau(T) \rightarrow 0$.

Note that if $Plim \  b(n(T),T) = 1$ or $0$,  then Bayesian-updated beliefs are correct with a probability going to 1. Thus, a threshold for learning is sharp in that if the number of chains of signals is of higher order, then the receiver learns the true state with a probability going to one, while if it is of
lower order, the receiver learns {\sl nothing}.

Recall that $M \equiv  1 - \mu_{01} - \mu_{10}$.

\begin{lemma}
	\label{multipath1}
Consider a learner getting signals from a network in $R(n(T), T)$ and suppose that $p_0=p_1 = p$.
Then $\frac{1}{p^TM^{2T}}$ is a threshold for learning.
\end{lemma}

All proofs are found in Appendix \ref{app:proofs}.

The threshold in Lemma \ref{multipath1} is sharp, and translates directly into a threshold for the average degree in the tree.
For instance,
suppose that the learner receives word-of-mouth signals
through a random tree generated by a Galton-Watson branching process  in which average degree distribution $F$ places weight 0 on degree 0 and has finite variance.\footnote%
{This condition ensures that the tree does not die out and so has at least some paths of depth $t$ with probability one. The analysis can be adapted to allow for extinction, but no new insight emerges.}
Then, one can easily show that
$Plim \ b(t) = 1$ or $0$ whenever $\mathbb{E}_{d\sim F}[d]>\frac{1}{pM^{2}}$ and (ii) $Plim \  b(t) = \theta$ whenever $\mathbb{E}_{d\sim F}[d]<\frac{1}{pM^{2}}$.

\subsection{The Impossibility of Learning with Uncertain Mutation Rates}\label{extremists}

Lemma \ref{multipath1} shows that learning is possible with sufficiently many sources - an amount growing exponentially in $T$ as a function of the mutation probabilities.
However, as we now show, the result above is dependent upon the rates of mutations being perfectly known.
In practice, there is likely to be some uncertainty about the mutation rates.
As we show next, slight uncertainty about the mutation rates {\sl completely}
precludes any learning in the limit.

\begin{proposition}
	\label{treeLearning3}
	Consider a learner getting signals from a network in $R(n(T), T)$, with dropping rates $p_1 = p_0=p$. If the learner does not know $\mu_{10}/\mu_{01}$, and has an atomless prior with convex support,
then for \emph{any} $n(T)$ the receiver learns nothing in the limit; i.e.,
the learner's posterior $b(n(T),T)$ converges in probability to $\theta$ as $T$ goes to infinity.
\end{proposition}

Proposition \ref{treeLearning3} shows that even a tiny amount of uncertainty about the relative mutation rates completely eliminates any possibility of limit learning, regardless of how many chains are observed.

The intuition is that, as $T$ grows, the fraction of 1 versus 0 signals converges to   $\frac{\mu_{01}}{\mu_{10}}$.

Learning comes from the fact that there is a bias away from $\frac{\mu_{01}}{\mu_{10}}$ in favor of the starting state, but that bias vanishes as $T$ grows.
If there are enough signals as a function of $T$ (so $n(T)$ grows fast enough), then that slight bias can be discerned - which is the result in Lemma \ref{multipath1}.
However, any uncertainty about $\frac{\mu_{01}}{\mu_{10}}$ completely swamps
the vanishing difference in the relative frequency of signals that reflects the starting
state.\footnote{Appendix \ref{app:impossibility_intuition}
illustrates the intuition by comparing short versus long (finite) chains
in an example where asymmetric mutations arise from idealogical
agents spreading disinformation. There, we also discuss the role of both uncertainty and
asymmetries in mutation in driving the impossibility result. In particular, uncertainty
over some other parameters of the environment need not preclude learning.}

\section{Overcoming Bias by Restricting Communication}

The `impossibility of learning' result of Proposition \ref{treeLearning3} provides an important intuition and benchmark.  However, it relies on the premise that the learner is distant from all of the original sources.
In practice, the learner is likely to have some sources closer than others.   This opens the possibility of partial learning.
We consider that issue in this section, considering networks in which information may originate from all nodes.

We study some learner's network which is a tree with depth $T$ in which each non-leaf node has in-degree $d$, with the learner at the root,
but in which any node can be a primary source. Information then reaches the learner with noise
depending on the distance
from each source.  In particular, we extend our model to allow any node to be an
original source with probability $r>0$.
Closer nodes have fewer mutations and fewer dropped messages, and are thus easier to learn from.
For simplicity, we consider cases in which the learner cannot tell the distance that information has traveled, which applies widely and
makes the tradeoff clear.

We start by considering a platform designer with an objective
of maximizing the number of messages that the learner receives subject to the
constraint that received messages are more likely to be true than false in both states.
This objective highlights the tradeoff between the volume and
veracity of socially-received messages.
It also ensures that naive learners (or Bayesians with a symmetric prior on mutation rates) who follow the majority of signals
will be correct with high probability, regardless of the starting state.

Expanding either breadth or depth increases the number of sources of information.
However, expanding depth affects the ratio of far to near nodes in a different way from expanding breadth.
Having more nearby sources relative to far away sources improves the chance that any given signal is correct.
Clearly, the optimum network from the learner's point of view is to have a huge $d$ and small $T$.
However, this requires that people have an unreasonably large number of
friends,\footnote{Since $r$ is likely to be low on many topics of importance, even people
with hundreds of friends may only have a few informed on any given topic.
For instance, if
only 1 in 10000 people is knowledgeable about some new medical treatment,
then in order to expect to have 30 informed friends, a person would have to have 300000 friends in total.}
and also that these many friends refrain from sharing information by word of mouth.

Given that people have a limited number of contacts
(and may often not know anyone directly informed about a specific issue),
some depth in people's networks is needed to generate information.
However, increasing depth also increases the noisiness of the signals that are received.
To analyze this tradeoff, we begin by identifying the largest depth $T$ from
which received messages are more likely to be true than false regardless of the state.
This threshold depth serves as a cap on chain length such that,
if chains longer than this are not permitted, then the majority of messages are expected to be true,
even from the longest distance and regardless of the starting state.

The interesting case is when mutation rates differ across states.  If $\mu_{01}=\mu_{10}$ then messages always are (slightly) more likely to match the starting state and there is no limit to the distance at
which truth is probabilistically preserved, even at a vanishing rate and even if these mutation rates are unknown.
The interesting challenge comes when $\mu_{01}\neq \mu_{10}$.
For instance, if $\mu_{01}< \mu_{10}$,
then as $T$ becomes very large any surviving signal is more likely to be a 0 than a 1, regardless of the starting state.  If the state is 0, this is not a problem, but if the state
is 1 then the mutations eventually make the majority of the messages false.  It is only for small enough $T$ that signals are more likely to still be correct when the true state is 1.
More generally, depending on which of $\mu_{01}$ and $ \mu_{10}$ is larger, one of the states becomes hard to learn from distant signals.

For simplicity, suppose that $0$ and $1$ messages survive with the same probability ($p_0=p_1 \equiv p >0$ -- we treat the heterogeneous case in the next section).
Moreover, suppose agents are in an infinite, directed $k$-regular tree.
That means each agent has an in-degree and out-degree of $k$, and any directed path between two nodes is unique.

Supposing that the state is 0, the frequency of true (0) to false (1) messages originating $t$ steps is,
\begin{equation}\frac{\mu_{10} + \mu_{01}M^t}{\mu_{01} - \mu_{01}M^t}.\label{ratio_0a}\end{equation}
Similarly, if the state is 1, the frequency of true (1) to false (0) messages originating $t$ steps is,
\begin{equation}\frac{\mu_{01} + \mu_{10}M^t}{\mu_{10} - \mu_{10}M^t}.\label{ratio_1a}\end{equation}

A platform designer who wants agents on average to hear more true messages than false, regardless of the state, needs to ensure that both (\ref{ratio_0a}) and (\ref{ratio_1a}) are greater than 1.
The cap on $T$ thus depends on which state it is relatively easier to mutate away from.

\begin{proposition}
\label{capT}
Messages that originate at a distance of less than
\[
T^*=\frac{\log(\frac{1}{2}) + \log(1 - \min\{\frac{\mu_{01}}{\mu_{10}}, \frac{\mu_{10}}{\mu_{01}}\})}{\log(M)},
\]
are more likely to be true than false, while messages that originate at a distance of greater
than $T^*$ are more likely to be false than true in the state from which it is easier mutate.
\end{proposition}

Since $M<1$, both the numerator and denominator are negative; so, $T^*$ is positive.
As $\mu_{01}$ and $\mu_{10}$ become more equal, then $T^*$ grows and approaches infinity in the limit of identical mutations.
At the other extreme, where one of the two mutation rates is 0 and the other is 1/2, then $T^*$ shrinks to 2.

If a planner knows a bound on  $ \min\{\frac{\mu_{01}}{\mu_{10}}, \frac{\mu_{10}}{\mu_{01}}\}$ -- which could be estimated from sampled data -- then
the cap can be calculated accordingly.

Generally, if $T$ can be capped, then increasing breadth improves learning; so, there is no reason to
limit breadth if depth can be capped.\footnote{There are some local non-monotonicities in
learning as one changes degree, as for instance
even numbers of signals can allow for ties among other things.  However, perfect learning is feasible in the limit as breadth is increased.}
However, capping $T$ requires following the life-cycle of a message, which can be infeasible in practice.
One reason is that it may be difficult or impossible to tell whether someone is mutating a previous
message she heard or originating a new message entirely.
Alternatively, a designer maintaining the privacy of users may not even observe
the messages being forwarded, which are encrypted on some platforms.
When depth cannot be restricted,
limiting breadth can offer substantial improvements in learning.
A designer can still lower the {\sl relative} number of long chains
by restricting the number of others to whom any given agent can forward messages.

To understand this, we examine a tree in which there is no cap on depth (i.e., an infinite tree),
and consider which cap on degree
guarantees more true than false messages in both states, in expectation.
Again, consider a tree in which each node has in-degree and out-degree $k>0$.
A cap on message forwards of $d\leq k$ limits each agent to passing along messages to at most $d$ of its out-neighbors.
We suppose that agents follow the cap by choosing their $d$ out-neighbors uniformly at random.   A cap of $d>k$ is not binding.
That also results in an effective average in-degree of $d$.

\begin{proposition}\label{platformpolicy}
If $\mu_{01} = \mu_{10} = \mu$, then the expected number of true messages exceeds the expected number of false messages regardless of degree and state. When
$\mu_{01} \neq \mu_{10} $, the expected number of true messages exceeds the expected number of false messages in both states if and only if
degree is less than
 $$d^* = \frac{1 - 2\max\{\mu_{01}, \mu_{10}\}}{pM}.$$
\end{proposition}

As $p$ becomes smaller, then fewer messages come from longer distances, and $d^*$ becomes larger.   Also,
if we increase the higher of the two mutation rates,
then $\frac{1 - 2\max\{\mu_{01}, \mu_{10}\}}{M} = \frac{1 - 2\max\{\mu_{01}, \mu_{10}\}}{1 - \mu_{01} - \mu_{10}}$
decreases and a tighter cap is needed.
In the extreme as $p$ approaches 1, so that messages are likely to travel far, then
 $d^*$ becomes less than 1.  At some point it becomes impossible to have a tree with average degree of
 more than 1.  If messages are almost always forwarded, then people need to be probabilistically
 capped at less than 1 forward to ensure that most messages are true.

Propositions \ref{capT} and \ref{platformpolicy} give tight bounds on when truthful messages are expected to outnumber false ones, as a function
of the depth and breadth of a learner's network.
Such caps do not necessarily maximize the probability that truthful messages outnumber false ones, as they analyze expected numbers of messages.
One may want to go further to maximize the probability that truthful messages outnumber false ones.   This involves
lower caps than the above bounds.
With lower caps, one ends up further restricting the number of messages, but also further increasing the relatively frequency of true to false messages in the harder state to learn.

\begin{figure}%[h!]
	\centering
	\includegraphics[scale=0.5]{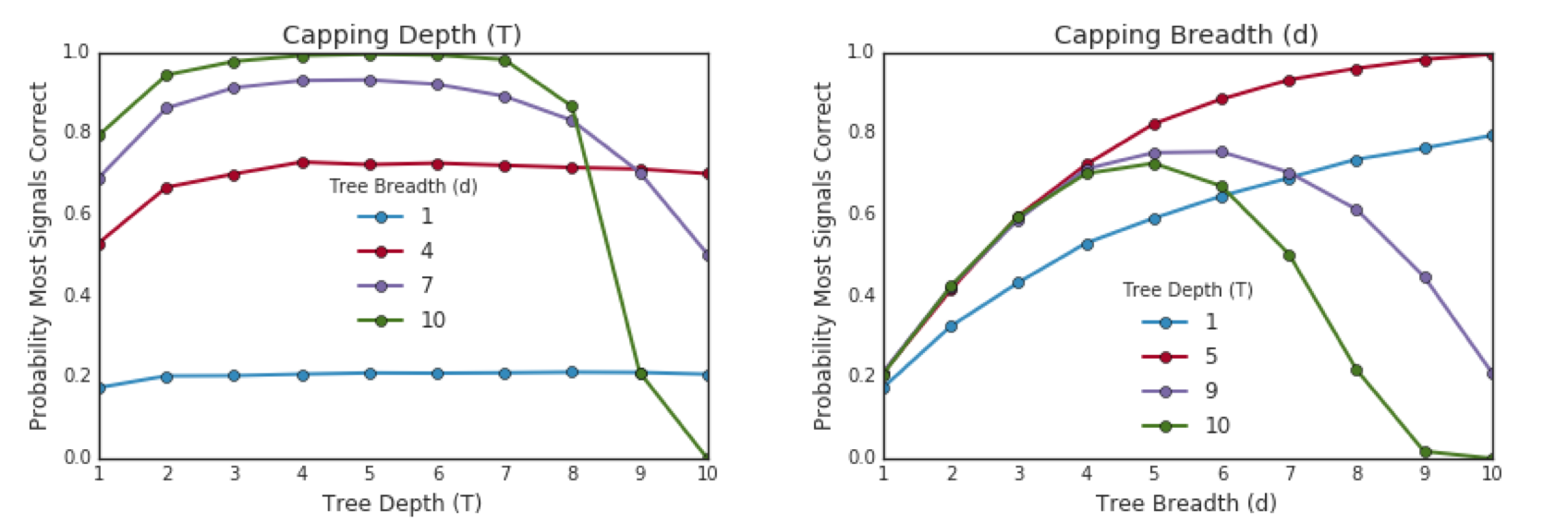}
	\caption{These graphs show how the {\sl probability} that most received messages are true in state 1 varies with the depth and breadth of the learner's social network. These values are simulated for the parameters $p_0=p_1=0.2$ and $\mu_{01}=0.1$ and $\mu_{10}=.03$.
}
	\label{fig:capping}
\end{figure}

The expression for the probability of having true messages exceed false ones appears to be intractable to optimize in a closed form; but the intuition is clear and so we illustrate it
in an example, for different combinations of $d,T$ as pictured in Figure \ref{fig:capping}.

For the parameter values behind the example pictured in Figure \ref{fig:capping},
Propositions \ref{capT} and \ref{platformpolicy} yield caps
of depth $T^*=7.5$ and breadth $d^*= 4.6$.
These graphs suggest that setting depth at $T=4$ or $T=5$ (for trees of any degree up to 10) maximizes the probability that more messages are true than false.
When setting breadth, the maximum is to set $d=5$ for trees of depth 9 or 10 (approaching the $d^*=4.6$ for the higher $T$), while for trees of
lower depth the optimum is to set degree to the maximum of 10.	

Our results above show platforms can reduce the frequency of
false communications by limiting either the depth,
or breadth of
people's information networks.
These policies enable agents to
partially learn the true state in the presence of asymmetric mutations,
by ensuring that
relatively more of an agent's received messages originate from nearby.

Most importantly, this policy improves learning without having to observe the content of messages. In many situations, such policies are attractive as they respect privacy.
Indeed they are the only feasible policies if messages sent over a platform are encrypted and unobservable to a platform designer. Even in cases where messages are observable, a platform designer who does not know the true state of the world cannot discern true messages from false ones. And even when she can, she might want to take a more neutral stance and stay out of the business of censoring messages based on their content.

Cutting down on ``fake news" has recently become an
expressed objective for social media platforms that have been blamed for
facilitating the spread of misinformation and disinformation. For example, the messaging
platform WhatsApp was criticized for allowing the spread of incendiary
rumors leading to mob killings in India and for serving as a vehicle of
misinformation and disinformation in 2018 Brazilian elections. In line with the results of
this section, WhatsApp placed a cap of 5 on the number of people to
whom a message can be forwarded in an attempt to curtail the spread of
false information (see \cite{hern_safi_2019}).

\section{An Extension:  Learning from Message Survival}\label{subsection_survival}
\label{surv}

So far, we focused on how (asymmetric) mutations hamper learning and how limiting networks restore it.
In this section, we consider how receivers learn when transmission failures can also depend on the content of the message being passed. This opens up the possibility that receivers learn not only from the content of received messages but also the frequency of communication on a given issue.

For example, it is often observed that statistically significant results are more likely to be shared and published than insignificant ones. If researchers in some academic field may know that many people are working on a certain topic, but do not hear of many results, then they may guess that most results were insignificant, even without paying attention to the content of the studies that were published and shared.

Formally, suppose $p_0 \neq p_1$, so that current signal content affects the likelihood that the signal is dropped along a word-of-mouth transmission chain.   Without loss of generality, we focus on the case in which $p_1 > p_0$, meaning that people are more likely to pass along signal 1 than 0. We return to the case of learning from sources that are the leaves of a tree in $R(n(T),T)$, as this crystalizes the comparison of learning from message content versus message survival.

Unlike the content of a single message, which becomes nearly meaningless as chains grow long
(due to mutation), the information conveyed by a single message's survival does not vanish in the
long-chain limit.\footnote{See Appendix \ref{app:single_chain} a more detailed comparison of leaning from
content versus survival with a single source.}
For intuition, suppose for a moment that only the first agent in each chain was biased in favor of
message 1, and other agents transmit with probability $\hat{p_1}$ regardless of signal content.
The likelihood of survival to $t$ is $p_1 \left( \hat{p_1} \right)^{t-1}$ if the first agent saw signal 1 or $p_0 \left( \hat{p_0} \right)^{t-1}$ if the first agent saw signal 0. Thus, the relative likelihood of survival equals $p_1/p_0 > 1$ (favoring signal 1) no matter how long the chain.  Moreover, biasing all agents in favor of transmitting message 1 further increases the relative likelihood of survival from state 1, since signal 1 is more likely to be received at each step along the chain when the true state is 1 rather than 0.

Before analyzing the full Bayesian model, we consider the challenge of learning for a receiver who only counts messages without checking what they say. As with learning from signal content only, the learner can discern the state from just signal frequency as long as transmission is more likely after one signal than the other ($p_0 \neq p_1$) and there are sufficiently many starting sources of information.

\begin{lemma}\label{learn_survival}
Consider a learner getting signals from a tree in  $R(n(T), T)$; suppose that $\mu_{01}=\mu_{10}=\mu \in ( 0,1/2]$ and $1>p_1>p_0>0$.\footnote{If $\mu=0$, then it is easy to check that the threshold is an expected degree of $1/p$, which is then the threshold for messages to survive conditional upon state $\omega=1$, which are the more likely to survive.}
	There exists $\lambda(T) = c + o(1)$ for some $c \in (0, 1)$, such that a threshold
for learning when conditioning only upon signal survival
is $$\frac{1}{(p_1 \lambda(T) + (1 - \lambda(T))p_0)^T}.$$
\end{lemma}

Given that messages mutate, the probability that any agent transmits a message lies somewhere between $p_1$ and $p_0$. Conditional on the initial message being 1, the overall probability that a message is transmitted all the way to the end of a length-$T$ chain must therefore take the form ${(p_1 \lambda + (1 - \lambda)p_0)^T}$ for some $\lambda \in (0,1)$. Only if the number of sequences $n(T)$ grows faster than this would a growing number of signals survive, conditional on the state being 1. The learner can then discern the state (perfectly in the limit) based on the actual number of signals that survive. As with the case of learning only from content, the threshold $n(T)$ grows exponentially in chain length.

\subsection{Full Bayesian Learning vs Learning Only from Survival or Only from Content}

In this section we compare how much better a Bayesian learner is at guessing the true state compared to someone who pays attention only to signal survival or only to signal content.

Without loss of generality, we focus on the case in which $p_1 \geq p_0$.

Suppose that mutations are symmetric and the learner has access to a
single chain of information resulting in signal $s_t \in \{0,1,\emptyset\}$.
We consider four different ways in which the learner might guess the state.

\begin{itemize}
\item A ``Bayesian agent,'' $B$, guesses the most likely state conditional on both signal survival and signal content.
\item A ``survival rule-of-thumb agent,''  $S$, guesses 1 if a signal is received and guesses 0 if
no signal is received.
\item A ``content rule-of-thumb agent,'' $C$, guesses 1 if signal 1 is received, 0 if signal 0 is received,
and guesses in favor of the prior if no message is received (flipping a coin if $\theta=1/2$).
\item A ``naive agent,''  $N$,  always guesses in favor of the prior.
\end{itemize}
$S,C,N$ are collectively referred to as ``limited learners'' since they make their guess based on
 less information than is available.

 If the limited learners have more than one source of
 information; i.e., learn from a network in $R(n(T), T)$, we can modify the descriptions of
 their behavior accordingly. ``C" guesses 1 whenever the fraction of 1 messages compared to
 0 messages is above a threshold, and ``S" guesses 1 whenever the number of messages that survive
 is above a threshold.  These thresholds are the conditional Bayesian ones, but these agents only
 consider one aspect of the available information.

\begin{proposition}
	\label{relativeLearning}
	Suppose that $1 \geq p_1 \geq p_0 \geq 0$ and $\mu_{01}=\mu_{10}=\mu \in [0,1/2]$.
	\begin{enumerate}
		\item (Single Source) The probability that a Bayesian agent is correct in predicting the state is at most
		$\frac{4}{3}$ higher
		than the best limited learner when $T=1$, and at most $\frac{3}{2}$ higher than the best limited learner for all $T>1$.\footnote{We conjecture that the bound is $\frac{4}{3}$ for any $T>1$.}
		Moreover, as $T$ grows, this upper bound converges to 1.
	
		\item (Multiple Sources) The threshold for learning is the same for $B$ as it is for the
		lowest of the thresholds corresponding to $C$ or $S$.
	\end{enumerate}

\end{proposition}

Proposition \ref{relativeLearning} implies that a belief-updating strategy that uses only message survival or only message content approximates one that uses all available information, no matter what the parameters or realized state. The result on multiple sources shows this approximation becomes exact in the limit, so full learning can be obtained from the more informative dimension of information.\footnote{This is obvious when $\mu=1/2$, in which case message content contains no information, or when $p=q$, in which case message survival contains no information.} The quantitative bounds along a single chain show that some naive learning strategy always performs well even for in the finite $T$ case.

\subsection{Uncertain Dropping Rates and Limiting Scope}

Next, let us consider what happens with uncertainty about mutation and relative dropping rates.
As $T$ grows, even with different dropping rates, only a vanishing fraction of chains survive.
 When $p_1\neq p_0$, changing $\mu_{01}/\mu_{10}$ slightly changes that fraction by orders of magnitude even though it will still be
vanishing.
Again, this crowds out the information about
the original state that can be gleaned from survival, which dies out
over the sequence.\footnote{In particular,
to see this (wlog) consider a case in which $1>p_1>p_0>0$.
Let $Z_{\mu_{01}/\mu_{10}}(T,s)$ be the probability that a signal survives $T$ periods conditional on starting out as a signal $s$, given $\mu_{01}/\mu_{10}$.
The key observation is that $Z_\pi(t,0)/ Z_{\mu_{01}/\mu_{10}-\varepsilon}(T,1)$ grows without bound as $T$ grows, for any $\varepsilon$.  Both probabilities are tending to 0, but
eventually they mix and the probabilities are strings of products of combinations of $p_1$s and $p_0$s, and tilting that combination one way or the other eventually accumulates arbitrarily in terms of
{\sl relative} probabilities as things are exponentiated.  Even a small shift in the fraction of mutations completely overturns the advantage of the starting state.
Then tiny uncertainty about $\mu_{01}/\mu_{10}$ introduces much larger swings in the survival rates than the starting states.}
Thus, the difficulty with learning  in the face of uncertain bias (Proposition \ref{treeLearning3}) is not overcome by considering survival when $p_1\neq p_0$.

Considering the case of $p_1\neq p_0$ changes the expressions for the caps on messages from Propositions \ref{capT} and \ref{platformpolicy}, but does not
affect the intuition or point.  What changes is that survival rates can make it more or less likely that true or false messages survive.
For instance, if the true state is 1, then increasing $p_1$ or decreasing $p_0$
leads to greater relative survival of true messages and a lower relative frequency of false messages.
That increases the $T^*$ cap, if 1 is the state that is harder to learn, while it decreases $T^*$ if 0
is harder to learn.

\section{Concluding Remarks}

We introduced a model of social learning via relayed signals in the presence of signal mutation, dropping,
and biasing.
We showed that learning is challenging in the presence of mutations and dropped messages:
complete learning requires an exponentially growing
number of original sources as the length of the chains over which information is relayed grows.
Once enough signals are present, simple rules of thumb approximate Bayesian updating.
However, learning from long chains depends on precise knowledge of mutation rates.
Uncertainty about mutation rates leads to a tradeoff between the size
of someone's network and their ability to learn.  In that context, capping
the depth of a network -- or the breadth if the depth cannot be capped -- improves learning.

These challenges can also motivate learners to seek out information from closer, trusted contacts,
and, if messages can
be traced in this way, to down-weight information from more distant sources.
To the extent that a platform can make it easy to trace the length of a chain,
and a message's history, that could also enhance learning.
However, tracing information backwards is difficult in many cases, especially for platforms that
allow messages to be encrypted or that do not check content out of respect for user privacy.
In such scenarios, caps on the breadth of users' word-of-mouth networks can be powerful tools,
and easily implemented.

\medskip

\bibliographystyle{aer}
\bibliography{noisycommunication}

\bigskip

\begin{appendices}
%\appendixpage
\section{Learning from Content Versus Survival Along a Single Chain}\label{app:single_chain}

Here, we characterize learning along a single chain of messages to build intuition for how learning purely from content ($p_0 = p_1$ case) compares to learning purely from survival ($\mu_{01} = \mu_{10}$ case).

First consider learning purely from content. Lemma \ref{miracle} provides some simple expressions that are useful and are straightforward Markov Chain formulas:

\begin{lemma}
	\label{miracle}
	Suppose that $p_0=p_1>0$ and consider any mutation rate $\mu_{01}, \mu_{10} \in (0,1/2)$. If the state is 0 and agent $t \geq 1$ receives a non-null signal, then the signal is 0 (matching the true state) with probability
	\begin{equation*}
	X_0(t) = \frac{\mu_{10} + \mu_{01}M^t}{\mu_{10} + \mu_{01}}.
	\end{equation*}
	If the state is 1 and agent $t \geq 1$ receives a non-null signal, the signal is 1 (matching the true state) with probability
	\begin{equation*}
	X_1(t) = \frac{\mu_{01} + \mu_{10}M^t}{\mu_{01} + \mu_{10}}.
	\end{equation*}
It follows that
\begin{equation*}
	X_0(t)- (1-X_1(t)) = M^t.
	\end{equation*}
As $t$ grows, regardless of the starting state, the probability that a surviving signal is a 0  and a 1, respectively are
\begin{equation*}
	\pi_0 = \frac{\mu_{10}}{\mu_{10} + \mu_{01}}    \ \ \  \mathrm{ and } \ \ \ \pi_1 = \frac{\mu_{01}}{\mu_{10} + \mu_{01}}.
	\end{equation*}
	Finally, if $\mu_{01} = \mu_{10} = \mu$, then the signal matches the true state with probability
	\begin{equation}\label{eqn_Xt}
	X(t) = \frac{1+M^t}{2}.
	\end{equation}
\end{lemma}

\bigskip

In contrast to the case where only message content is informative, when $p_0\neq p_1$ message survival continues to be informative about the true state of the world even as the chain of messages grows long. We formalize this statement and characterize how informative message survival can be in the following proposition. Let
\[ z  \equiv \frac{p_1}{p_0}\left( 1 + (1-2\mu)  \frac{(p_1-p_0)}{ p_0 + \mu(p_1-p_0)}\right).
\]
When $p_1>p_0$ and $0\geq \mu 1/2$,  $z> \frac{p_1}{p_0}>1$.  % is strictly greater than 1.

\begin{lemma}\label{facts}
	Suppose that
$1 \geq p_1 > p_0 > 0$ and  $\mu_{01}=\mu_{10}=\mu \in (0,1/2]$.\footnote%
	{If $\mu=0$ then $ \frac{Pr(s_t \neq \emptyset | \omega=1 )}{Pr(s_t \neq \emptyset | \omega=0 )}=(p_1/p_0)^t$, which
		diverges, and the problem becomes trivial.  Similarly, if $p_0=0$ then $Pr(s_t \neq \emptyset | \omega=0 ) =0$ and the problem becomes trivial. Note that here we do not require that $\mu<1/2$ since survival at the first step contains information, even if subsequent steps are completely random.  This contrasts with the case in which $p_1=p_0$, in which
		learning is precluded when $\mu=1/2$.}
	\begin{enumerate}
		\item The relative probability of message survival over a chain of
		length $t$ conditional on state 1 versus state 0 is uniformly bounded away from $\frac{p_1}{p_0}$:
		\begin{equation}\label{survivala}
		\frac{Pr(s_t \neq \emptyset | \omega=1 )}{Pr(s_t \neq \emptyset | \omega=0 )} \geq  z \geq  \frac{p_1}{p_0} \text{ for all } t \geq 1,
		\end{equation}
		with strict inequalities when $\mu<1/2$.
		\item The ratio in (\ref{survivala}) converges as chain-length grows:  $y\equiv \lim_{t\to \infty} \frac{Pr(s_t \neq \emptyset | \omega=1 )}{Pr(s_t \neq \emptyset | \omega=0 )}$ exists.
		\item Upon seeing a surviving message, the learner's updated belief  $Pr(\omega =1  | s_t \neq \emptyset)$ is uniformly bounded below by $\frac{\theta }{\theta + (1-\theta)/z} >  \theta$  and bounded above in the limit by $\frac{\theta }{\theta + (1-\theta)/y}<1$.\label{c1}
		\item In the limit, updating is entirely due to signal survival
		and not content: $\lim_{t \to \infty} Pr(\omega = 1 | s_t = 1) =
		\lim_{t \to \infty} Pr(\omega = 1 | s_t = 0) = \lim_{t\to \infty}
		Pr(\omega = 1 | s_t \neq \emptyset) $.\label{p2}
	\end{enumerate}
\end{lemma}

\section{Interpreting the Impossibility of Learning Result}\label{app:impossibility_intuition}

This appendix provides more intuition for Proposition \ref{treeLearning3}. Here we link the result to an example where some agents spread disinformation, illustrate how to interpret the impossibility result for networks of finite depth, and show why both uncertainty and potential asymmetry play a role in driving the result.

\medskip
\noindent
\textbf{Disinformation example.} Consider an example where the source of asymmetric mutations is the presence of evenly dispersed ideologues, who always pass on the message that they prefer. In particular, suppose that each agent is a ``1-ideologue'' with probability $\alpha \pi$ and a ``0-ideologue'' with probability $\alpha (1-\pi)$. $\alpha$ is the fraction of agents who are ideologues, and $\pi$ is the fraction of ideologues who favor message 1. For simplicity, suppose there are no unintentional mutations, $\mu = 0$, and no transmission failures, $p_0=p_1=1$.

Say that a signal is ``contaminated'' if any agent in the chain generating that signal is an ideologue. Each signal is uncontaminated with probability $(1-\alpha)^t$, which tends to 0. Each contaminated signal is determined by the last ideologue in the chain, and hence equals $1$ with probability $\pi$ and $0$ with probability $1-\pi$.

When the state is 1, a fraction $(1-\alpha)^t + \pi (1 - (1-\alpha)^t)$ of all signals equal 1, while if the true state is 0, a fraction $\pi(1 - (1-\alpha)^t)$ of signals equals 1.

When $\pi$ and $\alpha$ are exactly known, the vanishing difference $(1-\alpha)^t$ in expected signal content depending on the state allows the learner to discern the true state given enough signals.

However, with any uncertainty about $\pi$, the vanishing difference of $(1-\alpha)^t$ is completely obscured by the non-vanishing overall base level of $\pi (1 - (1-\alpha)^t)$, which varies nontrivially.
This logic holds no matter how many signals are received.

\medskip
\noindent
\textbf{Networks of finite depth.} The intuition that the receiver should be able to learn with sufficiently many signals only holds when those signals originate sufficiently nearby in the network.

Suppose $\pi$ in the preceding example has a very concentrated, atomless support, say $(\frac{1}{3} - \varepsilon, \frac{1}{3}+ \varepsilon)$ for some small $\varepsilon>0$.
Consider the case where signals all originate one step away from the receiver, i.e., $t=1$. If the state is 1, then in expectation, at least $(1-\alpha) + (\frac{1}{3}-\varepsilon) \alpha$ signals will be 1. If $t=0$, then in expectation, at most  $(\frac{1}{3}+\varepsilon) \alpha$ signals will equal 1. When $\alpha$ is sufficiently small, $(1-\alpha) + (\frac{1}{3}-\varepsilon) \alpha > (\frac{1}{3}+\varepsilon) \alpha$. So the receiver can discern between the two states and learn with sufficiently many signals.

Define the threshold $t(\alpha)$ implicitly by the condition $(1-\alpha)^{t(\alpha)} + (\frac{1}{3}-\varepsilon) (1 - (1-\alpha)^{t(\alpha)}) = (\frac{1}{3}+\varepsilon) (1 - (1-\alpha)^{t(\alpha)})$. So long as $t < t(\alpha)$, perfect learning remains possible given enough signals. On the other hand,  once $t > t(\alpha)$, the receiver cannot perfectly discern state 1 from state 0 even with arbitrarily many signals.

\medskip
\noindent
\textbf{Role of uncertainty and asymmetries in mutations.} Other forms of uncertainty can also hamper learning. For example, agents may have uncertainty about the average degree in the network (tree).
Again, as the observer is learning from a fractional difference in \textit{survival rates}, having a larger order uncertainty over the overall number of sequences obscures the finer information needed to infer the state.

Interestingly, without asymmetric mutation rates, uncertainty about the degree does not preclude learning from \textit{signal content}, since that is based on relative frequencies and not how many sequences survive.
Indeed, if mutations only took the form of random symmetric errors, uncertainty over the rate of mutation does not completely crowd out learning either.

\begin{proposition}
	\label{possibilityresult}
	Suppose the learner receives signals from $n(t)$ independent chains of length $t$, with dropping rates $p_0, p_1 \in [0, 1)$ mutation rate $\mu_{01} = \mu_{10} = \mu \in [0, \frac{1}{2})$. Moreover, suppose the learner knows $p_0$ and $p_1$ exactly, but only has a prior on $\mu$ with support bounded away from $\frac{1}{2}$. Then there exists $f(t)$ such that if $n(t)/f(t) \to \infty$, $Plim\ b(t) = 0$ or $1$.
\end{proposition}

The proof is simple consequence of Lemmas \ref{multipath1} and \ref{learn_survival}. There is a threshold for learning for every $\mu$ in the support of the agent's prior. Then having more signals than the threshold of learning for the largest such $\mu$ is sufficient for the agent to learn fully.

These cases illustrate that different types of noise and uncertainty may crowd out different forms of learning (i.e., learning from content alone, learning from survival alone, or learning from both).

\section{Proofs}\label{app:proofs}

\noindent {\bf Proof of  Lemma \ref{multipath1}:}

Consider the case in which the true state is 0, as the other case is analogous.

The expected number of 0 signals is $n(t) p^t X_0(t)$ while the expected number of  0 signals when the state is 1 is $n(t) p^t (1- X_1(t))$. The  difference in the expected number of 0 signals across states is
\[
D(t) \equiv n(t) p^t X_0(t)-n(t) p^t (1- X_1(t))= n(t) p^t M^t.
\]
If the standard deviation of the number of 0 signals in both states divided by $D(t)$ goes to 0, then by Chebychev's inequality, the probability of seeing more than $n(t) p^t (1- X_1(t)) + \frac{D(t)}{2}$ 0 signals when the state is 1 goes to zero. On the other hand, the probability of seeing fewer than this many 0 signals when the state is 0 goes to zero.

When the ratio of standard deviation to $D(t)$ goes to infinity, the likelihood ratio between both states is within $1 - \epsilon(t)$ on a $1 - \delta(t)$ measure of signals (say, according to the measure when the state is 0), where $\epsilon, \delta \to 0$. Therefore, there is no learning in the limit.

It is therefore enough to show that the ratio of standard deviation to $D(t)$ goes to either infinity or zero when $n(t)$ is above or below the threshold, respectively. Now, the standard deviation of the number of 0 signals in the 0 state divided by the amount above is
\[
\frac{ ( X_0(1-X_0)n(t) p^t)^{1/2}}{n(t) p^t M^t} =
\frac{ ( X_0(1-X_0))^{1/2}}{(n(t) p^t)^{1/2} M^t}.
\]

Note that the numerator converges to a constant, and so this expression either goes to 0 or infinity depending on whether
$(n(t) p^t)^{1/2} M^t$ goes to 0 or infinity, which depends on whether  $(n(t)pM^2)^t$ goes to 0 or infinity.

The expressions for all the other standard deviations and differences are analogous.\eproof

\medskip

\noindent {\bf Proof of  Proposition \ref{treeLearning3}:}

We give the proof for the case in which $p^t n(t) \rightarrow \infty$.
(With fewer paths there are even fewer signals from which to learn.)
The following lemma is straightforward (and so its proof is omitted) but it useful.

\begin{lemma}
	\label{lembinom}
	Consider a sequence of $k \leq m$ such that $m\rightarrow \infty$ and $\frac{k}{m}\rightarrow a$.
	The maximizer of $z^k (1-z)^{m-k}$ is $z(m,k)=\frac{k}{m}$, and
	\[
	\frac{ z(m,k)^k (1-z(m,k))^{m-k}}{z^k (1-z)^{m-k}} \rightarrow \infty
	\]
	for any $z\neq a$, the size of this ratio increases with the distance of $z$ from $a$ (as $z^a (1-z)^{1-a}$ is strictly concave).
	Moreover, for any atomless and continuous probability measure $G$ on $z$ that has connected support and includes $a$ in its interior
	\[
	\frac{\int_{a-\varepsilon}^{a+\varepsilon}  z^k (1-z)^{m-k}dG(z) }{\int_0^1  z^k (1-z)^{m-k} dG(z) } \rightarrow 1,
	\]
	for any $\varepsilon>0$.
\end{lemma}

Via a standard calculation for the limiting distribution for a two-state Markov chain, the steady state limit probability that the message is 1 is
\[
\frac{\mu_{01}}{\mu_{01} + \mu_{10}} \equiv \rho.
\]

The probability that some sequence ends in a 1 conditional on survival, $\rho$ and starting in state $\omega=1$ is
\[\rho + (1-\rho)M^t.\]

The probability that some sequence ends in a 1 conditional on survival, $\rho$ and starting in state $\omega=0$ is
\[\rho(1 - M^t).\]

Similar calculations provide probabilities of ending in a 0.

The chance of observing $k$ 1s, conditional on $m$ sequences reaching the receiver, on $\mu_{01}$ and on the starting state being $\omega=1$  is then
\[
P_{k,m,t,\rho}(1) =
\binom{m}{k} \left[\rho + (1-\rho)M^t\right]^k   \left[(1-\rho) (1-M^t)\right]^{m-k}.
\]
Then
the chance of observing $k$ 1s out of $m$ sequences that reach the receiver conditional on the starting state being $\omega=0$  is then
\[
P_{k,m,t,\rho}(0) =
\binom{m}{k} \left[\rho(1 - M^t)\right]^k   \left[(1-\rho) + \rho M^t\right]^{m-k}.
\]

First consider the case where $\rho$ is known, and suppose the state is 1 (the argument for the case where the state is 0 is analogous). As the number of signals grows large (keeping $t$ fixed), $\frac{k}{m-k} \to \frac{\mu_{01} + \mu_{10}M^t}{\mu_{10} - \mu_{10}M^t} \equiv a_{t,1}$ in probability. Suppose without loss of generality that $\mu_{01}\geq \mu_{10}$ so $a_{t,1}>1$ for all $t$.  Now, the Bayesian's posterior that the state is $\omega=1$ conditional upon seeing $k$ 1's out of $m$ sequences that reached the receiver
\[
\frac{\theta P_{k,m,t, \rho}(1)}{\theta P_{k,m,t, \rho}(1)+ (1-\theta) P_{k,m,t, \rho}(0)},
\]
Let $k_n$ and $m_n$ be the random number of 1's and messages received respectively with $n$ length $t$ chains. By Lemma \ref{lembinom}, $\frac{P_{k_n,m_n,t, \rho}(0)}{P_{k_n,m_n,t, \rho}(1)} \rightarrow 0$ in probability as the number of signals $n$ grow large, so it follows that
\[
\frac{\theta P_{k_n,m_n,t, \rho}(1)}{\theta P_{k_n,m_n,t, \rho}(1)+ (1-\theta) P_{k_n,m_n,t, \rho}(0)}\rightarrow 1.
\]

Therefore, since the agent can learn the true state with sufficiently many paths for any given $t$, it follows that the agent can learn the true state as $t\to \infty$ if $n(t)$ grows quickly enough.

Now we consider the case when $\rho$ is unknown but follows an atomless distribution $F$ with connected support. A Bayesian's posterior that the state is $\omega=1$ conditional upon seeing $k$ 1's out of $m$ sequences that reached the receiver is
\[
\frac{\theta \int_{\rho} P_{k,m,t,\rho }(1)dF(\rho)}{\theta \int_{\rho} P_{k,m,t,\rho}(1)dF(\rho)+ (1-\theta) \int_{\rho} P_{k,m,t,\rho }(0)dF(\rho)},
\]
and so if we can show that
$ \int_{\rho} P_{k,m,t,\rho }(1)dF(\rho) / \int_{\rho} P_{k,m,t,\rho}(0)dF(\rho)$ converges to one in probability,
then we conclude the proof.

Given a true ${\mu_{01}}^*, {\mu_{10}^*}$ such that $\frac{\mu_{10}^*}{\mu_{01}^*}$ in the interior of the support of $F$, the realized $k,m$ will be such that
$
\frac{k}{m-k} - \frac{\mu_{01}^* + \mu_{10}^*M^t}{\mu_{10}^* - \mu_{10}^*M^t} = \frac{k}{m-k} -\frac{\rho^* + (1-\rho^*)M^t}{(1-\rho^*) - (1-\rho^*)M^t}
$ converges to 0 in probability,
and $\frac{\rho^* + (1-\rho^*)M^t}{(1-\rho^*) - (1-\rho^*)M^t} \rightarrow \frac{\rho^* }{ 1-\rho^*} =a$.

By the first part of Lemma \ref{lembinom}, for any $k,m$,
$P_{k,m,t,\rho}(1)$ is maximized when $\rho$ equals $\rho(t, k, m, 1)$ such that
\[
\rho(t, k, m, 1) + (1 - \rho(t, k, m, 1))M^t = \frac{k}{m},
\]
and $P_{k,m,t,\rho}(0)$ is maximized when $\rho$ equals $\rho(t, k, m, 0)$ such that
\[\rho(t, k, m, 0) (1- M^t) = \frac{k}{m}.
\]

$\rho(t, k, m, 1)$ and $\rho(t, k, m, 0)$ converge to each other, and to $\rho^*$, as well in probability. It therefore follows from Lemma \ref{lembinom} that
\[
plim \ \frac{\int_{\rho} P_{k,m,t,\rho}(1)dF(\rho)}{ \int_{\rho} P_{k,m,t,\rho}(0)dF(\rho)} = plim \
\frac{\int_{\rho(t, k, m, 1)-\varepsilon}^{\rho(t, k, m, 1)+\varepsilon} P_{k,m,t,\rho}(1)dF(\rho)} {\int_{\rho(t, k, m, 0)-\varepsilon}^{\rho(t, k, m, 0)+\varepsilon} P_{k,m,t,\rho}(0)dF(\rho)}
\]
for any $\varepsilon > 0$.

Letting $[l,h]$ be the support of $\rho$, note that (a) $P_{k,m,t,\rho(t, k, m, 1)}(1) = P_{k,m,t,\rho(t, k, m, 0)}(0)$, and (b) that $\frac{d}{d\rho}P_{k,m,t,\rho}(1)$ evaluated at $\rho'$ and $\frac{d}{d\rho}P_{k,m,t,\rho}(0)$ evaluated at $\rho''$ are the same when $P_{k,m,t,\rho'}(1)=P_{k,m,t,\rho''}(0)$. It follows that

\[P_{k,m,t,\rho(t, k, m, 1)+\delta}(1) = P_{k,m,t,\rho(t, k, m, 0)+\delta}(0)\]
for any $\delta \in \mathbb{R}$ such that both $\rho(t, k, m, 1)+\delta$ and $\rho(t, k, m, 0)+\delta$ fall in $(l, h)$. In particular, if we let $\varepsilon_t = \frac{1}{2}\min\{\rho(t, k, m, 1) - l, h - \rho(t, k, m, 0)\}$ and if $\varepsilon_t > 0$, the intervals $[\rho(t, k, m, 1) - \varepsilon_t, \rho(t, k, m, 1) + \varepsilon_t]$ and $[\rho(t, k, m, 0) - \varepsilon_t, \rho(t, k, m, 0) + \varepsilon_t]$ strictly lie in $(l, h)$. So by the earlier observation,
\[\int_{\rho(t, k, m, 1)-\varepsilon_t}^{\rho(t, k, m, 1)+\varepsilon_t} P_{k,m,t,\rho}(1)dF(\rho) = \int_{\rho(t, k, m, 1)-\varepsilon_t}^{\rho(t, k, m, 1)+\varepsilon_t} P_{k,m,t,\rho}(0)dF(\rho)\]
Moreover $plim \ \varepsilon_t = \frac{1}{2}\min\{\rho^* - l, h - \rho^*\} > 0$. Therefore, by the continuous mapping theorem,
\[plim \ \frac{\int_{\rho} P_{k,m,t,\rho}(1)dF(\rho)}{ \int_{\rho} P_{k,m,t,\rho}(0)dF(\rho)} = plim \
\frac{\int_{\rho(t, k, m, 1)-\varepsilon_t}^{\rho(t, k, m, 1)+\varepsilon_t} P_{k,m,t,\rho}(1)dF(\rho)} {\int_{\mu_{01}(t, k, m, 0)-\varepsilon_t}^{\rho(t, k, m, 0)+\varepsilon_t} P_{k,m,t,\rho}(0)dF(\rho)}=1,\]
which concludes the proof.\eproof

\medskip

\noindent {\bf Proof of Lemma \ref{miracle}:}
We derive the expressions of $X_0,X_1$, which can also be deduced from standard Markov chain results, but it may be useful for the reader to see the derivation.
The proof is by induction. We give the proof for $X_0$, when the state is 0. The proof for $X_1$ is symmetric and the expression for $X$ is a special case.

First, note that if $t=1$ then this expression simplifies to $1-\mu_{01}$, which is exactly the probability that the signal has not mutated, and so this holds for $t=1$.

Then for the induction step, supposing that the claimed expression is correct for $t-1$, we show it is correct for $t$.

The probability of matching the true state at $t$ is the probability of not matching at $t-1$ times $\mu_{10}$ plus the probability of matching at $t-1$ times $1-\mu_{01}$,
which by the induction assumption can be written as
\begin{align*}
& \ \ \ \left[1-\frac{\mu_{10} + \mu_{01}M^{t-1}}{\mu_{10} + \mu_{01}}\right] \mu_{10}
+
\left[\frac{\mu_{10} + \mu_{01}M^{t-1}}{\mu_{10} + \mu_{01}}\right](1- \mu_{01})\\
&= \mu_{10} + \left[\frac{-\mu_{10}^2 - \mu_{01} \mu_{10} M^{t-1} + \mu_{10} - \mu_{10}\mu_{01} + \mu_{01}M^{t-1} - \mu_{01}^2M^{t-1}}{\mu_{10} + \mu_{01}}\right]\\
&= \mu_{10} + \left[\frac{-\mu_{10}^2 + \mu_{10} - \mu_{10}\mu_{01} + \mu_{01} M^{t-1}(1 -\mu_{10} - \mu_{01})}{\mu_{10} + \mu_{01}}\right]\\
&= \frac{\mu_{10}  + \mu_{01} M^t}{\mu_{10} + \mu_{01}}.
\end{align*}
as claimed.\eproof

\medskip

\noindent{\bf Proof of Proposition \ref{platformpolicy}:}

Suppose without loss of generality that $\mu_{01} \leq \mu_{10}$.

By Lemma \ref{miracle}, the expected number of true messages exceed the expected number of false messages when the state is $0$, since for every $t$, $X_0(t) = \frac{\mu_{10} + \mu_{01}M^t}{\mu_{10} + \mu_{01}} > \frac{\mu_{01} - \mu_{01}M^t}{\mu_{10} + \mu_{01}} = 1 - X_0(t)$.

It remains to be shown that the expected number of true messages also exceed the expected number of false messages when the state is 1. The expected number of true messages received is $\sum_{t=1}^{\infty} r(pd)^tX_1(t)$, and the expected number of false messages received is $\sum_{t=1}^{\infty} r(pd)^t(1-X_1(t))$.
So, we need to characterize the conditions under which
$$\sum_{t=1}^{\infty}r (pd)^tX_1(t)
>\sum_{t=1}^{\infty} r(pd)^t(1-X_1(t)).$$
Thus we need $\sum_{t=1}^{\infty} (pd)^t(2X_1(t) - 1) >0$; i.e., when
$$\sum_{t=1}^{\infty} (pd)^t(\frac{2\mu_{10}M^t + \mu_{01} - \mu_{10}}{\mu_{10} + \mu_{01}})  > 0.$$
If $\mu_{01} = \mu_{10}$, then this holds regardless of $d$, which establishes the first statement
of the proposition.  Next, suppose $\mu_{01} < \mu_{10}$.
Then it follows that
\begin{align*}
\sum_{t=1}^{\infty} (pd)^t(\frac{2\mu_{10}M^t + \mu_{01} - \mu_{10}}{\mu_{10} + \mu_{01}}) & > 0\\
\iff \sum_{t=1}^{\infty} (pMd)^t(2\mu_{10}) &> \sum_{t=1}^{\infty} (pd)^t(\mu_{10} - \mu_{01})\\
\iff \frac{pMd}{1 - pMd}(2\mu_{10}) &> \frac{pd}{1-pd}(\mu_{10} - \mu_{01})\\
\iff \frac{2\mu_{10} M}{1 - pMd} &> \frac{\mu_{10} - \mu_{01}}{1-pd}\\
\iff \frac{1 - pMd}{2\mu_{10} M} &< \frac{1-pd}{\mu_{10} - \mu_{01}}\\
\iff \frac{1}{2\mu_{10} M} - \frac{pd}{2\mu_{10}} &< \frac{1}{\mu_{10} - \mu_{01}} - \frac{pd}{\mu_{10} - \mu_{01}}\\
\iff \frac{1}{p}(\frac{1}{2\mu_{10} M} -  \frac{1}{\mu_{10} - \mu_{01}}) &< d(\frac{1}{2\mu_{10}} - \frac{1}{\mu_{10} - \mu_{01}})\\
\iff d &< \frac{1}{p}\frac{\frac{\mu_{10} - \mu_{01}}{M} -  2\mu_{10}}{{\mu_{10} - \mu_{01}} - 2\mu_{10}}\\
\iff d &< \frac{1}{p}\frac{2\mu_{10} + \frac{\mu_{01} - \mu_{10}}{M}}{\mu_{10} + \mu_{01}}\\
\iff d &< \frac{1}{pM}\frac{\mu_{10} + \mu_{01} - 2\mu_{10}(\mu_{10} + \mu_{01})}{\mu_{10} + \mu_{01}}\\
\iff d &< \frac{1 - 2\mu_{10}}{pM}.
\end{align*}
This is the condition in the proposition for $\mu_{01} < \mu_{10}$, which was without loss of generality, concluding the proof.\eproof

\medskip

\medskip

\noindent {\bf Proof of Lemma \ref{facts}, Part 1:}

For ease of notation,
let $P^t_{1S} \equiv Pr(s_t \neq \emptyset | \omega=1 )$
and $P^t_{0S} \equiv Pr(s_t \neq \emptyset | \omega=0 )$.
These are the probabilities of signal survival to time $t$ conditional on the first signal.

First we prove that
$\frac{P^t_{1S}}{ P^t_{0S}}\geq \frac{p_1}{p_0} $, with strict inequality when $\mu<1/2$ and $t>1$.

This is proven by induction.
First, $P^1_{1S} = p_1 > p_0=P^1_{0S} $.
Next, let us show that  $\frac{P^t_{1S}}{ P^t_{0S}}\geq \frac{p_1}{p_0}$ given that $P^{t-1}_{1S} >P^{t-1}_{0S} $.
Note that given the stationarity of the process, $P^{t-1}_{1S} = Pr(s_t \neq \emptyset | s_1=1)$ and $P^{t-1}_{0S} = Pr(s_t \neq \emptyset | s_1=0)$, and
then we can write\footnote{The starting state $s_0$ is 1 in this calculation and so then there is a probability $p$ that the signal survives to the first period,
	and then the calculation inside the $\left[ \cdot\right]$ handles the two possible values of the first period signal and then the probability
	the signal survives to $t$ if it has made it to the first period in the two possible values it could have in the first period.}
The first part
$$P^t_{1S} = p_1 \left[(1-\mu) Pr(s_t \neq \emptyset | s_1=1) + \mu  Pr(s_t \neq \emptyset | s_1=0)\right],$$
and so then it follows that
$$P^t_{1S} = p_1 (1-\mu) P^{t-1}_{1S} + p_1\mu  P^{t-1}_{0S}.$$
Then by the inductive step ($P^{t-1}_{1S} >P^{t-1}_{0S} $) and so it follows that
$$P^t_{1S} \geq p_1 (1-\mu) P^{t-1}_{0S} + p_1\mu  P^{t-1}_{1S},$$
with strict inequality when $\mu< 1/2$ and $t>1$.
Similarly,
$$P^t_{0S} =p_0 (1-\mu) P^{t-1}_{0S} + p_0\mu  P^{t-1}_{1S}.$$
Therefore
$$\frac{P^t_{1S}}{ P^t_{0S}} \geq \left(\frac{p_1}{p_0}\right)\frac{ (1-\mu) P^{t-1}_{0S} + \mu  P^{t-1}_{1S}}{  (1-\mu) P^{t-1}_{0S} + \mu  P^{t-1}_{1S}}=\frac{p_1}{p_0},$$
with strict inequality when $\mu< 1/2$ and $t>1$,
as claimed.

Now we complete the proof of the first part of the lemma.
Note that (from above)
$$\frac{P^t_{1S}}{ P^t_{0S}} = \left(\frac{p_1}{p_0}\right)\frac{ (1-\mu) P^{t-1}_{1S} + \mu  P^{t-1}_{0S}}{  (1-\mu) P^{t-1}_{0S} + \mu  P^{t-1}_{1S}}.$$
Therefore,
$$\frac{P^t_{1S}}{ P^t_{0S}} = \left(\frac{p_1}{p_0}\right)\left(\frac{ (1-\mu) P^{t-1}_{0S} + \mu  P^{t-1}_{1S} + (1-2\mu)(P^{t-1}_{1S}-P^{t-1}_{0S})}{  (1-\mu) P^{t-1}_{0S} + \mu  P^{t-1}_{1S}}\right),$$
and then since $p_1>p_0$ and  $\frac{P^{t-1}_{1S}}{ P^{t-1}_{0S}}\geq \frac{p_1}{p_0} $, with strict inequality when $\mu< 1/2$ and $t>1$, it follows that
$$\frac{P^t_{1S}}{ P^t_{0S}}=
\left(\frac{p_1}{p_0}\right)\left(1+ (1-2\mu)\frac{P^{t-1}_{1S}-P^{t-1}_{0S}}{ P^{t-1}_{0S}+ \mu(P^{t-1}_{1S}-P^{t-1}_{0S}})\right) \geq \frac{p_1}{p_0}\left( 1 +  (1-2\mu) \frac{(p_1-p_0)}{ p_0 + \mu(p-p_0)}\right),$$
with strict inequality when $\mu< 1/2$ and $t>1$ (and it directly follows that this expression ($z$) is strictly larger than $p_1/p_0$ when $\mu<1/2$),
as claimed.\eproof

\medskip

The following result is useful in the proofs of the remaining parts of Lemma \ref{facts}.

\begin{lemma}\label{ordering_probs}
	Fix $\theta \in (0, 1), \mu \in (0, 1/2], 0 < p_0 \leq p_1 \leq 1$. For all $t>0$,
	\[Pr(s_t = 1 | \omega = 1) \geq Pr(s_t = 0 | \omega = 1).\]
	Moreover,  either there exists $T$ large enough such that
	\[
	Pr(s_t = 1 | \omega = 0) \geq Pr(s_t = 0 | \omega = 0)
	\text{ \ for all \ } \text{ for all } t\geq T,
	\]
	or
	\[
	Pr(s_t = 1 | \omega = 0) < Pr(s_t = 0 | \omega = 0) \text{ for all } t.
	\]
	Finally, the sequence
	\[ \frac{Pr(s_t = 1 | \omega = 1)}{\min\{Pr(s_t = 1 | \omega = 0), Pr(s_t = 0 | \omega = 0)\}}\]
	is bounded above.
\end{lemma}

\noindent {\bf Proof of Lemma \ref{ordering_probs}:}

The first claim is proven by induction:

Since $\mu \leq 1/2$, $Pr(s_1 = 1 | \omega = 1) \geq Pr(s_1 = 0 | \omega = 1)$. Suppose $Pr(s_t = 1 | \omega = 1) \geq Pr(s_t = 0 | \omega = 1)$.
Note that,
\begin{align*}
Pr(s_{t+1} = 1 | \omega = 1) &= p_1 (1 - \mu) Pr(s_t = 1 | \omega = 1) + p_0 \mu Pr(s_t = 0 | \omega = 1) \\
Pr(s_{t+1} = 0 | \omega = 1) &= p_1 \mu Pr(s_t = 1 | \omega = 1) + p_0(1 -  \mu) Pr(s_t = 0 | \omega = 1).
\end{align*}
The result then follows from the inductive hypothesis and the facts that $p \geq q$ and $\mu \leq 1/2$.

Next, to show the second claim in the lemma, note that
\begin{align*}
Pr(s_{t+1} = 1 | \omega = 0) &= p_1 (1 - \mu) Pr(s_t = 1 | \omega = 0) + p_0 \mu Pr(s_t = 0 | \omega = 0) \\
Pr(s_{t+1} = 0 | \omega = 0) &= p_1 \mu Pr(s_t = 1 | \omega = 0) + p_0(1 -  \mu) Pr(s_t = 0 | \omega = 0).
\end{align*}
Then if $Pr(s_t = 1 | \omega = 0) \geq Pr(s_t = 0 | \omega = 0)$ for some $t=T$, the same will hold for all $t>T$ by a similar inductive proof.
Otherwise $Pr(s_t = 1 | \omega = 0) < Pr(s_t = 0 | \omega = 0)$ for all $t$, and then the result holds directly.

Finally, we show the third part of the claim.
By the second part of this lemma, there are two cases to consider. If  $Pr(s_t = 1 | \omega = 0) < Pr(s_t = 0 | \omega = 0)$ for all $t$. Then
\[
\frac{Pr(s_t = 1 | \omega = 1)}{\min\{Pr(s_t = 1 | \omega = 0), Pr(s_t = 0 | \omega = 0)\}} =
\frac{Pr(s_t = 1 | \omega = 1)}{Pr(s_t = 1 | \omega = 0)}
\]
If instead there is a $T$ such that for all $t \geq T$,  $Pr(s_t = 1 | \omega = 0) \geq Pr(s_t = 0 | \omega = 0)$, then
\begin{align*}
\frac{Pr(s_t = 1 | \omega = 1)}{\min\{Pr(s_t = 1 | \omega = 0), Pr(s_t = 0 | \omega = 0)\}} &=
\frac{Pr(s_t = 1 | \omega = 1)}{Pr(s_t = 0 | \omega = 0)} \\
&= \frac{p(1-\mu) Pr(s_{t-1} = 1 | \omega=1) + q\mu Pr(s_{t-1}=0 | \omega=1)}{p\mu Pr(s_{t-1} = 1 | \omega=0) + q(1-\mu) Pr(s_{t-1}=0 | \omega=0)} \\
&\leq \frac{(p_1(1-\mu) + p_0\mu) Pr(s_{t-1} = 1 | \omega=1)}{p_1\mu Pr(s_{t-1} = 1 | \omega=0) + p_0(1-\mu) Pr(s_{t-1}=0 | \omega=0)} \\
&< \frac{p_1(1-\mu) + p_0\mu}{p_1\mu}\frac{Pr(s_{t-1} = 1 | \omega = 1)}{Pr(s_{t-1} = 1 | \omega = 0)},
\end{align*}
where the second to last inequality uses the first part of this lemma. We can therefore handle both cases simultaneously by showing that the sequence $\frac{Pr(s_{t} = 1 | \omega = 1)}{Pr(s_{t} = 1 | \omega = 0)}$ is bounded above.

To that end, note that
\begin{align*}
Pr(s_t =1|\omega=0) &\geq Pr(s_t = 1 | \omega=0, s_1 = 1) Pr(s_1 = 1|\omega=0) \\
&= Pr(s_{t-1} = 1 | \omega=1) p_0\mu.
\end{align*}
So,
\[
\frac{Pr(s_{t} = 1 | \omega = 1)}{Pr(s_{t} = 1 | \omega = 0)} \leq \frac{Pr(s_{t} = 1 | \omega = 1)}{Pr(s_{t-1} = 1 | \omega = 1)}\frac{1}{p_0\mu}.
\]
It then suffices to show that ${Pr(s_{t} = 1 | \omega = 1)}\leq {Pr(s_{t-1} = 1 | \omega = 1)}$, since then
from above
\[
\frac{Pr(s_{t} = 1 | \omega = 1)}{Pr(s_{t} = 1 | \omega = 0)} \leq \frac{1}{p_0\mu},
\]
which is finite given that $p_0>0$ and $\mu>0$.
To see that  ${Pr(s_{t} = 1 | \omega = 1)}\leq {Pr(s_{t-1} = 1 | \omega = 1)}$,
\begin{align*}
Pr(s_{t} = 1 | \omega = 1)
&= p_1 (1 - \mu) Pr(s_{t} = 1 | s_1 = 1) + p_1\mu Pr(s_{t} = 1 |s_1 = 0)\\
&= p_1 (1 - \mu) Pr(s_{t-1} = 1 | \omega = 1) + p_1\mu Pr(s_{t-1} = 1 | \omega = 0)\\
&\leq  p_1 (1 - \mu) Pr(s_{t-1} = 1 | \omega = 1) + p_1 \mu Pr(s_{t-1} = 1 | \omega = 1)\\
&=  p_1  Pr(s_{t-1} = 1 | \omega = 1),
\end{align*}
where the inequality follows from the first part of the lemma, establishing the claim.\eproof

\medskip

\noindent {\bf Proof of Lemma \ref{facts}, Part 2:}

We show that $\lim_{t \to \infty}\frac{P^t_{1S}}{ P^t_{0S}} = \lim_{t \to \infty} \frac{p_1 Pr(s_{t-1} = 1 | \omega=1) + p_0 Pr(s_{t-1} = 0 | \omega=1)}{p_1 Pr(s_{t-1} = 1 | \omega=0) + p_0 Pr(s_{t-1} = 0 | \omega=0)}$ exists.

The sequence is bounded above by the first and last part of Lemma \ref{ordering_probs}:
it is bounded above by either $\frac{Pr(s_{t} = 1 | \omega=1)}{Pr(s_{t} = 1 | \omega=0)}$
or $\frac{Pr(s_{t} = 1 | \omega=1)}{Pr(s_{t} = 0 | \omega=0)}$, both of which are bounded above.
Furthermore, the sequence is bounded below by the first part of Lemma \ref{facts}.

To complete the proof that the limit exists, we show that the sequence is monotone. For this, we will start by writing, $r_t$, the $t^{th}$ term in the sequence, as $\frac{Pr(s_{t-1} = 1 | \omega=1) + \ell_1 Pr(s_{t-1} = 0 | \omega=1)}{ Pr(s_{t-1} = 1 | \omega=0) + \ell_1 Pr(s_{t-1} = 0 | \omega=0)}$, where $\ell_1 = p_0/p_1$. Now the $t+1^{st}$ is
\begin{align*}
r_{t+1} &= \frac{Pr(s_{t} = 1 | \omega=1) + \ell_1 Pr(s_{t} = 0 | \omega=1)}{ Pr(s_{t} = 1 | \omega=0) + \ell_1 Pr(s_{t} = 0 | \omega=0)} \\
&= \frac{(p_1(1-\mu) + \ell_1 p_1 \mu) Pr(s_{t} = 1 | s_1=1)  +  (p_0 \mu + \ell_1 p_0(1 - \mu))Pr(s_{t} = 0 | s_1=1)}{(p_1(1-\mu) +  \ell_1 p_1\mu) Pr(s_{t} = 1 | s_1=0) + (p_0\mu + \ell_1 p_0(1 - \mu))Pr(s_{t} = 0 | s_1=0)}\\
&=  \frac{Pr(s_{t-1} = 1 | \omega=1) + \ell_2 Pr(s_{t-1} = 0 | \omega=1)}{Pr(s_{t-1} = 1 | \omega=0) + \ell_2 Pr(s_{t-1} = 0 | \omega=0)},
\end{align*}
where $\ell_2 = \frac{p_0}{p_1}\frac{\mu + l (1 - \mu)}{(1-\mu) + l \mu}$. Consider the sequence $\ell_t$, where $\ell_{t+1} = \frac{p_0}{p_1}\frac{\mu + \ell_t (1 - \mu)}{(1-\mu) + \ell_t \mu}$ and $\ell_1 = \frac{p_0}{p_1}$.
Note that $\ell_t$ is non-decreasing in $t$ given that $\mu\leq 1/2$ and it is strictly increasing when $\mu<1/2$.
Iterating
on the above logic
\[r_t = \frac{Pr(s_{1} = 1 | \omega=1) + \ell_{t-1} Pr(s_{1} = 0 | \omega=1)}{ Pr(s_{1} = 1 | \omega=0) + \ell_{t-1} Pr(s_{1} = 0 | \omega=0)}.\]
To see that $r_t$ is monotone in $t$, note that the sign of the derivative of $r_t$ with respect to $\ell_t$ only
depends on the sign of
$Pr(s_{1} = 0 | \omega=1)Pr(s_{1} = 1 | \omega=0) - Pr(s_{1} = 1 | \omega=1)Pr(s_{1} = 0 | \omega=0)$),
and so it is monotone given the monotonicity of $\ell_t$ in $t$.\eproof

\medskip

\noindent {\bf Proof of Lemma \ref{facts}, Part 3:}

That $Pr(\omega = 1| s_t \neq \emptyset) \geq \frac{\theta z}{1 + \theta(z - 1)}$ for any $t>1$, with strict inequality when $\mu<1/2$, follows
from Part 1 and Bayes rule (and it is evident from the proof that this lower bound is not
tight). Therefore, it remains to show that $\lim_{t \to \infty} Pr(\omega = 1| s_t \neq \emptyset)$
exists, a step which is deferred to the proof of Part 4.

The fact that $\lim_{t\to \infty} Pr(\omega =1  | s_t \neq \emptyset) = \frac{\theta }{\theta + (1-\theta)/y}<1$ follows from Part 2 and Bayes' Rule.\eproof

\medskip

\noindent {\bf Proof of Lemma \ref{facts}, Part 4:}

It suffices to show that $\lim_{t \to \infty} Pr(\omega = 1| s_t = 1) = \lim_{t \to \infty} Pr(\omega = 1| s_t = 0)$, as this implies that $\lim_{t \to \infty} Pr(\omega = 1| s_t \neq \emptyset)$ exists and has the same value. This limiting equality between posterior distributions can equivalently be expressed in terms of likelihood ratios:
\begin{align*}
\lim_{t \to \infty} \frac{Pr(s_t=1 | \omega=0)}{Pr(s_t=1 | \omega=1)} &= \lim_{t \to \infty} \frac{Pr(s_t=0 | \omega=0)}{Pr(s_t=0 | \omega=1)} \\
\iff \lim_{t \to \infty} \frac{Pr(s_t=1 | s_t \neq \emptyset, \omega=0)}{Pr(s_t=1 | s_t \neq \emptyset, \omega=1)} &= \lim_{t \to \infty} \frac{Pr(s_t=0 | s_t \neq \emptyset, \omega=0)}{Pr(s_t=0 | s_t \neq \emptyset, \omega=1)}.\addtocounter{equation}{1}\tag{\theequation} \label{nts}
\end{align*}
We show that
\begin{equation}\label{nts1}
\lim_{t \to \infty} Pr(s_t=1 | s_t \neq \emptyset, \omega=0) = \lim_{t \to \infty} Pr(s_t=1 | s_t \neq \emptyset, \omega=1),
\end{equation}
since this implies that both sides of equation \ref{nts} are equal to 1.\footnote{Subtract each side of equation \ref{nts1} from 1 before taking ratios to see that the right side of equation \ref{nts} is also 1.}

Denote by $S$ a sequence of signals that evolve according to our process, starting with $S_0=1$ and $S'$ another (independent) sequence of signals with $S'_0=0$. Let $\tau = \min \{t | S'_t = 1\}$, where $\tau = \infty$ if $S'$ is dropped at some step before mutating to signal $1$, or if $S'_t = 0$ for all $t$.

In this notation, equation \ref{nts1} can equivalently be expressed as:  $\lim_{t \to \infty} Pr(S_t = 1 | S_t \neq \emptyset) = \lim_{t \to \infty} Pr(S'_t = 1 | S'_t \neq \emptyset)$.
Note the following relationship between the two independent paths:\footnote{Note that if $\tau>t$, then the probability that $S'_t=1$ is 0.}
\begin{align*}
Pr(S'_t = 1 | S'_t \neq \emptyset) &= \sum_{i=1}^{t} Pr(S'_t = 1 | S'_t \neq \emptyset, \tau = i) Pr(\tau = i | S'_t \neq \emptyset) \\
&=  \sum_{i=1}^{t} Pr(S_{t-i} = 1 | S_{t-i} \neq \emptyset) Pr(\tau = i |S'_t \neq \emptyset) \\
&\equiv \left( \sum_{i=1}^{t} Pr(S_{t-i} = 1 | S_{t-i} \neq \emptyset) w_i^t \right)
\addtocounter{equation}{1}\tag{\theequation} \label{sumexpanded},
\end{align*}
where $w_i^t = Pr(\tau = i |S'_t \neq \emptyset)$.

The result then follows from the following three claims, to be proved:
\begin{enumerate}
	\item For any $\varepsilon > 0$ and positive integer $k$, for all sufficiently large $t$, $\sum_{i=t-k}^{t} w^t_i < \varepsilon$.\label{fact1}
	\item $\lim_{t \to \infty} Pr(S_t = 1 | S_t \neq \emptyset)$ exists.\label{fact2}
	\item  $\sum_{i =1}^t w_i^t + w_{\infty}^t = 1$. Moreover, $w_{\infty}^t \to 0$ as $t \to \infty$, i.e., the probability that the signal never mutated conditional on survival to $t$ goes to 0 as $t$ grows.\label{fact3}
\end{enumerate}

To see that these claims imply the result, note that by claim \ref{fact1}, most of the weight falls on the first $t-k$ terms of the sum in equation  \ref{sumexpanded} for large enough $t$.  By claim \ref{fact2}, for a large enough $k$ (growing slower than $t$), these first $t-k$ terms will be close to  $\lim_{t \to \infty} Pr(S_t = 1 | S_t \neq \emptyset)$, and therefore by claim \ref{fact3} the limiting weighted sum of these terms converges to this value as well.

Claim \ref{fact3} is clear, so we prove the other two.

First we prove claim \ref{fact1}. Note that $p_0^i(1-\mu)^{i-1}\mu$ is the probability of survival with no mutation through $i-1$ and then survival with mutation at $t = i$, i.e., $Pr(\tau = i) = p_0^i(1-\mu)^{i-1}\mu$. Second, let $m_i$ be number of mutations through time $i$. Obviously, $Pr(S'_i \neq \emptyset) > Pr(S'_i \neq \emptyset$ and $m_i = 1)$. Third, if survival were always at rate $p_0$, then $Pr(S'_i \neq \emptyset$ and $m_i = 1) = i p_0^i(1-\mu)^{i-1}\mu$. However, since survival likelihood immediately after the first mutation, $p$, is strictly higher than $p_0$ and mutations sometimes occur (note, we assume $\mu > 0$), $Pr(S'_i \neq \emptyset$ and $m_i = 1) > i p_0^i(1-\mu)^{i-1}\mu$. Putting these observations together, we have
\begin{equation}\label{tau_ineq}
\lim_{i \to \infty} \frac{Pr(\tau = i)}{Pr(S'_i \neq \emptyset)} < \lim_{i \to \infty} \frac{p_0^i(1-\mu)^{i-1}\mu}{i p_0^i(1-\mu)^{i-1}\mu}  =\lim_{i \to \infty} \frac{1}{i} = 0,
\end{equation}
where, as noted earlier, the inequality arises from replacing $Pr(S'_i \neq \emptyset)$ with a lower bound on the probability of exactly one mutation occurring over the course of the first $i$ periods, and all the ways this could happen, and then
noting that $p_0<p_1$.
Now
\begin{align*}
Pr(\tau = i | S'_t \neq \emptyset) &= \frac{Pr(S'_t \neq \emptyset | \tau = i)Pr(\tau = i)}{Pr(S'_t \neq \emptyset)} \\
&= \frac{Pr(S_{t-i} \neq \emptyset)Pr(\tau = i)}{Pr(S'_t \neq \emptyset)} \\
&= \frac{Pr(S_{t-i} \neq \emptyset)Pr(\tau = i)}{Pr(S'_{t-i} = 1)Pr(S_i \neq \emptyset) + Pr(S'_{t-i} = 0)Pr(S'_i \neq \emptyset)} \\
&< \frac{Pr(S_{t-i} \neq \emptyset)}{Pr(S'_{t-i} \neq \emptyset)} \frac{Pr(\tau = i)}{Pr(S'_i \neq \emptyset)},
\end{align*}
where the inequality follows from the fact that $Pr(S_i \neq \emptyset) > Pr(S'_i \neq \emptyset)$, by Lemma \ref{facts}, Part 1. $\frac{Pr(S_{t-i} \neq \emptyset)}{Pr(S'_{t-i} \neq \emptyset)}$ is bounded by Lemma \ref{facts} Part 2 (as it has a limit), and $\frac{Pr(\tau = i)}{Pr(S'_i \neq \emptyset)}$ can be made arbitrarily small for large enough $i$ by equation \ref{tau_ineq}.
Thus, for any $\delta$ and $k$ we can find large enough $t$ for which $w_i^t< \delta$ for $i>t-k$.
Choosing $\delta=\varepsilon/k$ establishes claim \ref{fact1}.

Finally, we prove claim \ref{fact2}. The probability distribution of $S_t$ is given by $e_1^\prime A^t$, where $$ A =
\begin{bmatrix}
p_1(1 - \mu) & p_1\mu & 0 \\
p_0\mu & p_0(1-\mu) & 0\\
0 & 0 & 1
\end{bmatrix}$$ is the Markov transition matrix for $S$. Let $B$ be the principal $2\times 2$ submatrix of $A$.
By the partitioned matrix multiplication formula, $Pr(S_t = 1 | S_t \neq \emptyset) = \frac{e_1^\prime B^t e_1}{e_1^\prime B^t \mathbf{1}}.$ Since $B$ is strictly positive, the Perron-Frobenius theorem implies that this expression converges to the first entry of eigenvector corresponding to the largest eigenvalue of $B$.\eproof

\medskip

\noindent {\bf Proof of Lemma \ref{learn_survival}:}

$\lim_{t \to \infty}\frac{P^t_{0S}}{ P^t_{1S} } = r$ for some $r<1$, by Lemma \ref{facts}. Let $r_t$ be the $t^{th}$ term in the sequence.

Let $m(t)$ be the number of surviving signals.
By Chernoff bounds, it follows that
\[
Pr ( m(t) > n(t) P^t_{1s} (1+r_t)/2 | \omega = 1) \rightarrow 1
\]
and
\[
Pr ( m(t) < n(t) P^t_{1s} (1+r_t)/2 | \omega = 0) \rightarrow 1
\]
provided that $n(t) P^t_{1s} \rightarrow \infty$.
Given this separation, it is easy to the check that if $n(t) P^t_{1s} \rightarrow \infty$, the beliefs will converge to 0 or 1 in probability.

Next, note that if $n(t) P^t_{1s} \rightarrow 0$, then the expected number of surviving signals in either state is 0, and that happens with the probability going to 1 by Chebychev, and so
there is no learning.   So, the threshold is $1/ P^t_{1s}$.

Note that survival lies between $1/p_1^t$ and $1/p_0^t$ and so
$$1/ P^t_{1s} = \frac{1}{(p-1 \lambda(t) + (1 - \lambda(t))p_0)^t}.$$
The fact that $\lambda(t)$ converges to some $\lambda$ then follows
since this is a Markov chain and the probability that it survives in any
given period (the third state with $s_t=\emptyset$ is absorbing) converges
to a steady state distribution, which in this case lies between $p_1$ and $p_0$.\eproof

\medskip

The next lemma is useful in the  proof of Proposition \ref{relativeLearning}.

Let $P_1^t$ ($P_0^t$) denote the Bayesian posterior probability that the state is 1 conditional upon a signal being received at time $t$ and being 1 (0).  Similarly, let $P_\emptyset^t$ ($P_S^t$) denote the  Bayesian posterior probability that the state is 1 conditional upon no signal (some signal) being received at time $t$.

\begin{lemma}\label{relativeLearningUtilityLemma}
	If $p_1>p_0$, then $P_1^t\geq P_0^t$  and  $P_1^t\geq P_\emptyset^t$.
\end{lemma}

\noindent{\bf Proof of Lemma \ref{relativeLearningUtilityLemma}}

Let $s^t$ denote the state of the signal at period $t$. That $P_1^t\geq P_0^t$ holds when $t=1$ is easy to check from Bayes rule, given that $p_1> p_0$ and
$\mu\leq 1/2$.
Now suppose $P_1^t\geq P_0^t$ for some $t$. Then by the law of total probability, it follows that
\begin{align*}
P_1^{t+1} &= Pr(s^t = 0 | s^{t+1} = 1) P_0^t + Pr(s^t=1 | s^{t+1} = 1) P_1^t \\
&= \frac{p_0\mu Pr(s^t = 0)}{p_0\mu Pr(s^t = 0) + p_1 (1- \mu) Pr(s^t=1)} P_0^t + \frac{p_1 (1- \mu) Pr(s^t=1)}{p_0\mu Pr(s^t = 0) + p_1 (1- \mu) Pr(s^t=1)} P_1^t
\end{align*}
Similarly,
\[P_0^{t+1} = \frac{p_0(1-\mu) Pr(s^t = 0)}{p_0(1-\mu) Pr(s^t = 0) + p_1 \mu Pr(s^t=1)} P_0^t + \frac{p_1 \mu Pr(s^t=1)}{p_0(1-\mu) Pr(s^t = 0) + p_1 \mu Pr(s^t=1)} P_1^t\]
Since $P_1^t \geq P_0^t$ by the inductive hypothesis, it suffices to show that
\[\frac{p_1 (1- \mu) Pr(s^t=1)}{p_0\mu Pr(s^t = 0) + p_1 (1- \mu) Pr(s^t=1)} \geq \frac{p_1 \mu Pr(s^t=1)}{p_0(1-\mu) Pr(s^t = 0) + p_1 \mu Pr(s^t=1)}\]
i.e., that
\[\frac{1}{1 + \frac{p_0}{p_1}\frac{Pr(s^t = 0)}{Pr(s^t=1)}\frac{\mu}{1-\mu}} \geq \frac{1}{1 + \frac{p_0}{p_1}\frac{Pr(s^t = 0)}{Pr(s^t=1)}\frac{1-\mu}{\mu}}\]
which follows, since $\mu \leq 1-\mu$.

\ \ \ \ To see that $P_1^t \geq P_{\emptyset}^t$, note that it suffices to prove that $P_S^t \geq P_{\emptyset}^t$, since $P_S^t$ is a convex combination of $P_1^t$ and $P_0^t$, and we just proved $P_1^t \geq P_0^t$. Now the statement follows directly from part 1 of Proposition \ref{facts}.\eproof

\medskip
\noindent {\bf Proof of Proposition \ref{relativeLearning} (Single Source):}

First, note that we can focus on the case in which $p_1\neq p_0$ as otherwise there is nothing to be learned from signal survival, and agent $C$ does as well as $B$.
Without loss of generality we take $p_1>p_0$.   Similarly, if $\mu=1/2$, then all learning is from survival and $S$ does as well as $B$, and so we can take $\mu<1/2$.

Note that by Lemma \ref{relativeLearningUtilityLemma},  $P^t_1\geq P^t_0$  and  $P^t_1\geq P^t_\emptyset$. In order for $B$ to do strictly better in expectation than the other agents, it must be that $P^t_1>1/2$ and at least one of $P^t_0$ and $P^t_\emptyset$ are less than 1/2.
To see this note that if all three are on the same side of 1/2, then they must lie on the same side as the prior.\footnote{They cover three disjoint events whose union is all possibilities, and so the overall probability of a 1 is a convex combination of these conditionals, and so it is impossible to have them all weakly and some strictly greater (or all weakly and some strictly less) than the prior.}   If $\theta\neq 1/2$ then $N$ gets the same payoff as $B$.
If $\theta=1/2$,  then for all three to lie on the same side of the prior it must be that $p_1=p_0$,
in which case there is nothing learned from survival and $C$ does as well as $B$ in expectation.

Thus, $P^t_1>1/2$ and at least one of $P^t_0$ and $P^t_\emptyset$ are less than 1/2.
If it is just $P^t_\emptyset$ that is less than 1/2, then $S$ guesses the same as $B$ (or equivalently in expected payoff terms).
Thus, we need $P^t_0<1/2$ to have a difference.

If is just $P^t_0$ that is less than 1/2, then $C$ guesses the same as $B$ except if $\theta \leq 1/2$.  But for such a $\theta$, it must be that $P^t_\emptyset \leq 1/2$ and so $C$ guesses as well as $B$.

So, consider the case in which $P^t_1>1/2$ and $P^t_0<1/2$ and $P^t_\emptyset <  1/2$. For $C$ to guess differently than $B$, it must be that $\theta\geq 1/2$.

We can compute the expected payoff's for the three most relevant agents for this remaining case (we ignore $N$ now, since in these conditions it is dominated by one of the others) for a given $(p_1, p_0, \mu,\theta)$ satisfying the above constraints.

Letting $U_B, U_C, U_S$ be the expected payoffs of agents $B, C$\footnote{$C$ has expected payoff
	$U_C = Pr(s_t=1)P^t_1 + Pr(s_t=0)(1-P^t_0)
	+(1-Pr(s_t=1)-Pr(s_t=0)) \left(I_{\theta > 1/2} P^t_\emptyset
	+ I_{\theta = 1/2}  1/2 \right).$
	The expression in the main text is obtained by noting that the worst ratio for this compared to $B$ will be in cases for which $\theta>1/2$} and $S$ respectively, it follows that
\begin{align*}
U_B &= Pr(s_t=1)P^t_1 + Pr(s_t=0)(1-P^t_0)  + (1-Pr(s_t=1)-Pr(s_t=0)) (1-P^t_\emptyset) \\
U_C &= Pr(s_t=1)P^t_1 + Pr(s_t=0)(1-P^t_0) + (1-Pr(s_t=1)-Pr(s_t=0)) P^t_\emptyset \\
U_S &= Pr(s_t=1)P^t_1 +  Pr(s_t=0)P^t_0 + (1-Pr(s_t=1)-Pr(s_t=0))(1-P^t_\emptyset)
\end{align*}
First, note that if $p_0<1$ and $\mu>0$, then as $t\rightarrow \infty$, then  $Pr(s_t=\emptyset) \rightarrow 1$ and $P_\emptyset \rightarrow 1/2$, in which case the ratio of $B$ to either of these goes to 1. If $p_0<1$ and $\mu=0$, then $B$ does as well as $S$ for every $t$. If $p_1=p_0=1$, then $B$ does as well as $C$ for every $t$. These facts together establish the last claim in the proposition that as $t \to \infty$, the ratio $\frac{U_B}{\max\{U_S, U_C\}} \to 1$.

That the ratio is bounded above by 3/2 can be seen as follows.
Since $\theta\geq 1/2$ and $p_1> p_0$, it follows that
\[
Pr(s_t=1) \geq  Pr(s_t=0),   \ \ \   P^t_1 \geq  (1-P^t_0),  {\rm  \ \  and \  so \ \ }  Pr(s_t=1)P^t_1 \geq  Pr(s_t=0)(1-P^t_0).
\]
Then if $Pr(s_t=0)(1-P^t_0)\leq   (1-Pr(s_t=1)-Pr(s_t=0))(1-P^t_\emptyset)$ it follows that $U_S\geq  U_B 2/3$.
If $Pr(s_t=0)(1-P^t_0)\geq   (1-Pr(s_t=1)-Pr(s_t=0))(1-P^t_\emptyset)$ then it follows that $U_C\geq  U_B 2/3$.

To complete the proof, we compute
\[\max_{p_1, p_0, \theta, \mu \in [0, 1]} \frac{U_B}{\max\{U_S, U_C\}}.\]
for $t=1$.   We can rewrite the payoffs of agents $B$, $S$ and $C$ in the case $P_1^1 > 1/2$ and $P_0^1< 1/2$ and $P_{\emptyset}^1 < 1/2$ as follows:
\begin{align*}
U_B &= \theta p_1 (1 - \mu) + (1 - \theta) (1 - p_0 \mu) \\
U_C &= \theta (1 - p_1 \mu) + (1- \theta) p_0 (1 - \mu) \\
U_S &= \theta p_1                + (1 - \theta) (1 - p_0)
\end{align*}
where
\begin{align}
\theta p_1 \mu &\leq p_0 (1 - \theta) (1 - \mu) \label{bbeatss} \\
\theta (1 - p_1) &\leq (1 - \theta) (1 - p_0) \label{bbeatsc}\\
\theta &\geq 1/2 \label{thetarange}\\
\mu &\leq 1/2 \label{murange}\\
p_1 &\geq p_0 \label{survivalcond}\\
p_1, p_0, \mu, \theta &\in [0, 1] \label{paramrange}.
\end{align}

\textbf{Case 1}:  $U_S \leq U_C$.

This condition can be rewritten as
\begin{equation}\label{cbeatss}\theta(p_1\mu + (p_1-1)) \leq (1 - \theta)(p_0(1 - \mu) + (p_0 - 1))\end{equation}
The program with this additional constraint can be written as
\begin{align*}
\max_{p_1, p_0, \theta, \mu \text{ satisfy \ref{bbeatss}-\ref{cbeatss}}} \frac{U_B}{U_C} &\equiv \max_{p_1, p_0, \theta, \mu \text{ satisfy \ref{bbeatss}-\ref{cbeatss}}} \frac{\theta p_1 (1 - \mu) + (1 - \theta) (1 - p_0 \mu)}{\theta (1 - p_1 \mu) + (1- \theta) p_0 (1 - \mu)}\\
&= \max_{p_1, p_0, \theta, \mu \text{ satisfy \ref{bbeatss}-\ref{cbeatss}}} \frac{\theta p_1 + (1 - \theta) - \mu(\theta p_1 + (1 - \theta) p_0)}{\theta + (1 - \theta) p_0 - \mu (\theta p_1 + (1 - \theta) p_0)}\\
&\leq \max_{p_1, p_0, \theta, \mu \text{ satisfy \ref{bbeatss}-\ref{cbeatss}}} \frac{\theta + (1 - \theta) 2p_0 - 2\mu (\theta p_1 + (1 - \theta) p_0)}{\theta + (1 - \theta) p_0 - \mu (\theta p_1 + (1 - \theta) p_0)}
\end{align*}
where the inequality is from rearranging constraint \ref{cbeatss}, as $\theta p_1 + (1 - \theta) \leq \theta + (1-\theta)2p_0 - \mu(\theta p_1 + (1- \theta) p_0)$, and plugging this into the numerator. It is easily verified that the above ratio is decreasing in $\mu$ for any values of the remaining parameters \footnote{$\frac{d}{dx} \frac{A - 2x}{B - x} \leq 0$ if $A \leq 2 B$ and $A, B > 0$.}. Moreover, reducing $\mu$ to 0 only relaxes constraints \ref{bbeatss}, \ref{murange} and \ref{cbeatss}, and leaves the other constraints unaffected. Therefore,
$$
\max_{p_1, p_0, \theta, \mu \text{ satisfy \ref{bbeatss}-\ref{cbeatss}}} \frac{U_B}{U_C} \leq  \max_{p_1, p_0, \theta \text{ satisfy \ref{bbeatsc}-\ref{cbeatss}}} \frac{\theta + (1 - \theta) 2p_0}{\theta + (1 - \theta) p_0}
$$

It is clear that smaller values of $\theta$ increase this ratio, and by constraint \ref{thetarange}, the smallest value of $\theta$ is $\frac{1}{2}$. But while reducing $\theta$ down to $\frac{1}{2}$ for given $p_1$ and $p_0$ relaxes constraint \ref{bbeatsc}, doing so may violate constraint \ref{cbeatss}. We therefore separately consider the cases where either \ref{cbeatss} or \ref{thetarange} bind, since at least one of them must at the optimum.

\textbf{ Subcase 1:} \ref{cbeatss} is satisfied with equality, i.e., $\theta(1 - p_1) = (1 - \theta) (1 - 2p_0)$. Plugging this in, the objective then becomes $2\frac{1 + \theta(p_1-1)}{1 + \theta(p_1-1) + \theta}$, which is decreasing in $\theta$, so it is optimal to set $\theta=\frac{1}{2}$. The objective is then $2 \frac{1+p_1}{2+p_1} \leq 4/3$.
Note that at $p_1=1, p_0=\frac{1}{2}, \theta=\frac{1}{2}, \mu=0$, $\frac{U_B}{U_C} = \frac{4}{3}$, so this upper bound is tight.

\textbf{ Subcase 2:} $\theta=1/2$. Then $$\frac{U_B}{U_C} = \frac{p_1 + 1 - \mu(p_1+p_0)}{p_0+1 - \mu(p_1+p_0)},$$
which is weakly increasing in $\mu$ by constraint \ref{survivalcond}. Constraint \ref{cbeatss} can be rearranged to be $$\mu \leq \frac{2p_0 - p_1}{p_1 + p_0},$$
which, first, implies that $$\frac{U_B}{U_C} \leq \frac{2(p_1-p_0) + 1}{(p_1-p_0)+1},$$ and second, along with the condition that $\mu \geq 0$, implies that $$p_0 \geq \frac{p_1}{2}.$$
Since $\frac{2(p_1-p_0) + 1}{(p_1-p_0)+1}$ is decreasing in $p_0$, this expression is maximized under the given constraints when $p_0 = \frac{p_1}{2}$. Therefore, $\frac{U_B}{U_C} \leq \frac{p_1 + 1}{\frac{p_1}{2}+1}$, which is maximized when $p_1=1$ and equals $4/3$.

\textbf{Case 2:}  $U_S \geq U_C$.
The new constraint is
\begin{equation}\label{sbeatsc}\theta(p_1\mu + (p_1-1)) \geq (1 - \theta)(p_0(1 - \mu) + (p_0 - 1)),\end{equation}
and the relevant maximization program is

$$\max_{p_1, p_0, \theta, \mu \text{ satisfy \ref{bbeatsc}-\ref{paramrange}, \ref{sbeatsc}}} \frac{U_B}{U_S} \equiv \max_{p_1, p_0, \theta, \mu \text{ satisfy \ref{bbeatsc}-\ref{paramrange}, \ref{sbeatsc}}} \frac{\theta p_1 (1 - \mu) + (1 - \theta) (1 - p_0 \mu)}{\theta p_1 + (1- \theta) (1 - p_0)}.$$

Notice that the ratio $\frac{\theta p_1 (1 - \mu) + (1 - \theta) (1 - p_0 \mu)}{\theta p_1 + (1- \theta) (1 - p_0)}$ is linear and decreasing in $\mu$, and the constraints are linear in $\mu$ as well. Constraint \ref{bbeatss} only places an upper bound on $\mu$, so it is not relevant in pinning down this value at the optimum. On the other hand, constraint \ref{sbeatsc}, which can be rewritten as
\[2p_0(1-\theta) - p_1\theta - (1-2\theta) \leq (p_1\theta + p_0(1-\theta))\mu\]
and the constraint that $\mu \geq 0$ are relevant. There are two cases:

\textbf{Subcase 1: $2p_0(1-\theta) - p_1\theta - (1-2\theta) \geq 0$, $\mu=\frac{2p_0(1-\theta) - p_1\theta - (1-2\theta)}{p_1\theta + p_0(1-\theta)}$}.

Then
\begin{align*}
\frac{U_B}{U_S} &= \frac{\theta p_1 + (1 - \theta) - \mu (\theta p_1 + (1 - \theta) p_0)}{\theta p_1 + (1 - \theta) - (1 - \theta) p_0} \\
&=  \frac{\theta p_1 + (1 - \theta) - 2(1-\theta) p_0+ p_1\theta + (1 - 2\theta)}{\theta p_1 + (1 - \theta) - (1 - \theta) p_0} \\
&= \frac{2 \theta p_1 + (2 - 3\theta) - 2(1-\theta)p_0 }{\theta p_1 + (1 - \theta) - (1 - \theta) p_0} \\
&= 2 \frac{ \theta p_1 + (1 - \frac{3}{2}\theta) - (1-\theta)p_0 }{\theta p_1 + (1 - \theta) - (1 - \theta) p_0}
\end{align*}
Clearly, the ratio is decreasing in $\theta$, and moreover, decreasing $\theta$ only relaxes constraints \ref{bbeatss} and \ref{bbeatsc}. Therefore, constraint \ref{thetarange} binds and $\theta = \frac{1}{2}$ at the optimum, so

\[\frac{U_B}{U_S} = 2 \frac{p_1 + \frac{1}{2} - p_0}{p_1 + 1 - p_0}\]

Since $\mu = \frac{2p_0 - p_1}{p_1 + p_0}$ at $\theta=\frac{1}{2}$, constraints \ref{bbeatss} and \ref{bbeatsc} reduce to just $p_1 \geq p_0$. Since the ratio is increasing in $p_1 - p_0$, the only binding constraint is that $\mu \geq 0$, i.e., $2p_0 \geq p_1$. Therefore at the optimum, $p_1=1$, $p_0 = \frac{1}{2}$, $\mu=0$, $\theta=\frac{1}{2}$, and $\frac{U_B}{U_S} = \frac{4}{3}$.

\textbf{Subcase 2: $2p_0(1-\theta) - p_1\theta - (1-2\theta) \leq 0$, $\mu=0$.} In this case, the problem reduces to
\[\max_{p_1, p_0, \theta, \text{ satisfy \ref{bbeatsc},\ref{thetarange},\ref{survivalcond}, \ref{paramrange}, \ref{sbeatsc}}} \frac{\theta p_1  + (1 - \theta)}{\theta p_1 + (1- \theta) (1 - p_0)},\]
which is decreasing in $\theta$. Now
\begin{align*}
2p_0(1-\theta) - p_1\theta - (1-2\theta) &\leq 0 \\
\iff \theta(2 - 2p_0 - p_1) &\leq 1 - 2p_0
\end{align*}
Suppose $2 - 2p_0 - p_1 < 0$. Since $1 - 2p_0 \leq 1 - 2p_0 +(1-p_1) = 2 - 2p_0 - p_1 < 0$, it follows that $\theta (2 - 2p_0 - p_1) \geq \theta (1 - 2p_0) \geq 1 - 2p_0$.
Therefore, the only way to satisfy the constraint is if $p_1=1$ and $\theta=1$, in which case $\frac{U_B}{U_S} = 1$.

If $2 - 2p_0 - p_1 \geq 0$, then $\theta = \frac{1}{2}$
at the optimum, and so the constraint in this sub-subcase becomes $2p_0 \leq p_1$, while the objective function
is $\frac{p_1+1}{p_1 + 1 - p_0}$.
This constraint binds at the optimum and again the optimal value is $\frac{4}{3}$ at $p_1=1$
and $p_0=\frac{1}{2}$.\eproof

\medskip

The next lemma is useful in the proof of the second part of Proposition \ref{relativeLearning}.

\begin{lemma}
	\label{Bayes}
	Let $p_0=p_1$ and suppose a Bayesian agent with prior $\theta$ on the state being $1$ observes $m$ signals of which $k$ are $1$s.
Then the agent's posterior that the state is 1 is
	\[
	\frac{\theta X_1(t)^{k}(1 - X_1(t))^{m-k}}{\theta X_1(t)^{k}(1 - X_1(t))^{m-k} + (1-\theta) (1-X_0(t))^{k}X_0(t)^{m-k}}. %= \frac{\theta(1 + \delta^t)^k}{\theta(1 + \delta^t)^k + (1-\theta)(1 - \delta^t)^k}.
	\]
\end{lemma}
The proof is direct and omitted.

\noindent{\bf Proof of Proposition \ref{relativeLearning} (Multiple Sources):}

We proceed by cases for different values of the parameters.  We concentrate on situations in which $\mu<1/2$ since if $\mu=1/2$ then content is completely uninformative and  the result is direct.

\textbf{Case 1}: $\mu=0$. Suppose without loss of generality that $p_0 \leq p_1$. Any signal that reaches the agent is perfectly informative of the state, so a threshold for learning for agents B and C is the threshold for at least one signal to survive, which  (following the logic of the proofs above) is $\frac{1}{p_1^t}$.

\textbf{Case 2}: $p_1=p_0$ and $\mu>0$. By Proposition \ref{multipath1}, the threshold for learning for agent B is $\frac{1}{p_1^t (1 - 2\mu)^{2t}}$.  In this case there is no information from signal survival, and by Lemma \ref{Bayes}, agent B's posterior is the same as agent C's  posterior. Therefore, agent $C$ has the same threshold for learning as B.

\textbf{Case 3}:  $p_0 \neq p_1$ and $\mu>0$.
Without loss of generality let $p_1>p_0$. Then $\tau(t) = \frac{1}{P^t_{1S}}$ is a threshold for learning for an agent conditioning only on signal survival, as shown in the proof of Lemma \ref{learn_survival}.
Let $b(t)$ denote the beliefs of agent B after observing the outcome of $n(t)$ original sources of information sent along chains of depth $t$.
Since agent $B$ conditions on survival and signal content, $plim \ b(t) \to 1$ or 0 whenever $n(t)/\tau(t) \to \infty$.
When  $n(t)/\tau(t) \to 0$, then the probability of even a single signal surviving to reach the
agent approaches 0. This holds regardless of the starting state by Lemma \ref{facts} part 2, so   $plim \  b(t) \to \theta$. Therefore, agent B and S have the same thresholds for learning in this case.\footnote{Strictly speaking, we only showed that they share a common threshold, but it is easy to see that being a threshold for learning for B, for S or for neither partitions the space of functions on $\mathbb{N} \to \mathbb{N}$.}.\eproof

\end{appendices}

\end{document}